\documentclass[sn-standardnature]{sn-jnl}

\usepackage[numbers]{natbib}
\usepackage{mathtools}
\usepackage{physics}
\usepackage{url}
\usepackage{hyperref}
\usepackage{algorithm}
\usepackage{algpseudocode}
\usepackage{tabularx}
\usepackage{makecell}
\usepackage{subfig}
\usepackage{hyphenat}
\usepackage{microtype}
\usepackage{epstopdf}
\usepackage{ragged2e}
\usepackage{dutchcal}
\usepackage{xr}
\usepackage{xcolor}

\jyear{2022}%

\theoremstyle{thmstyleone}%
\theoremstyle{thmstyletwo}%

\theoremstyle{thmstylethree}%

\makeatletter

\newcommand*{\addFileDependency}[1]{
\typeout{(#1)}
\@addtofilelist{#1}
\IfFileExists{#1}{}{\typeout{No file #1.}}
}\makeatother

\newcommand*{\myexternaldocument}[1]{%
\externaldocument{#1}%
\addFileDependency{#1.tex}%
\addFileDependency{#1.aux}%
}

\myexternaldocument{supplygacs}

\begin{document}

\title [ ]{Effective implementation of $l_0$-Regularised Compressed Sensing with Chaotic-Amplitude-Controlled Coherent Ising Machines}


\author*[1]{\fnm{Mastiyage Don Sudeera Hasaranga Gunathilaka} }\email{mastiyage.s.aa@m.titech.ac.jp}

\author[2]{\fnm{Satoshi} \sur{Kako}}

\author[2]{\fnm{Yoshitaka} \sur{Inui}}
\author[1,4]{\fnm{Kazushi} \sur{Mimura}}

\author[5]{\fnm{Masato} \sur{Okada}}

\author[2,3]{\fnm{Yoshihisa} \sur{Yamamoto}}

\author[1]{\fnm{Toru} \sur{Aonishi}}

\affil*[1]{\orgdiv{School of Computing}, \orgname{Tokyo Institute of Technology}, \orgaddress{\street{Yokohama}, \city{Kanagawa}, \country{Japan}}}

\affil[2]{\orgdiv{Physics and Informatics Laboratories}, \orgname{NTT Research Inc.}, \orgaddress{\street{940 Stewart Dr}, \city{Sunnyvale, CA}, \postcode{94085}, \country{USA}}}

\affil[3]{\orgdiv{E. L. Ginzton Laboratory}, \orgname{Stanford University}, \orgaddress{\street{Stanford}, \city{CA}, \postcode{94305}, \country{USA}}}

\affil[4]{\orgdiv{Graduate School of Information Sciences}, \orgname{Hiroshima City University}, \orgaddress{ \city{Hiroshima}, \country{Japan}}}

\affil[5]{\orgdiv{Graduate School of Frontier Sciences}, \orgname{The University of Tokyo}, \orgaddress{\street{Kashiwa}, \city{Chiba}, \country{Japan}}}


\abstract{Coherent Ising Machine (CIM) is a network of optical parametric oscillators that can solve large-scale combinatorial optimisation problems by finding the ground state of an Ising Hamiltonian. As a practical application of CIM, Aonishi \textit{et al}., proposed a quantum-classical hybrid system to solve optimisation problems of $l_0$-regularisation-based compressed sensing. In the hybrid system, the CIM was an open-loop system without an amplitude control feedback loop. In this case, the hybrid system is enhanced by using a closed-loop CIM to achieve chaotic behaviour around the target amplitude, which would enable escaping from local minima in the energy landscape. Both artificial and magnetic resonance image data were used for the testing of our proposed closed-loop system. Compared with the open-loop system, the results of this study demonstrate an improved degree of accuracy and a wider range of effectiveness.}


\keywords{Coherent Ising machine, Compressed Sensing, Bayesian inference, Magnetic Resonance Imaging, Quantum–Classical hybrid
system, LASSO, Zeeman term, Chaotic-Amplitude Control, Gaussian-approximation, Closed-loop system, Combinatorial optimisation}



\maketitle

\section{Introduction}\label{intro}

Compressed sensing (CS) is a method of reconstructing a high-dimensional signal or image based on highly downsampled measurements. 

There has been considerable interest in it across a wide range of fields and applications. Such as in the field of astronomy, a possible way to transmit data to Earth from spacecraft \cite{CSastro} has been attempted. And there are proposed methods with CS on astronomical image compression and in compression on remotely sensed data \cite{CSastro2, CSastro3, CSastro4} as well. And in radar technologies for the reconstruction of the target image CS has been used \cite{CSsig2}. On the other hand in the medical field using embedded compression using CS to improve energy efficiency in Electrocardiogram (ECG) machines has been proposed \cite{CSbiosig}.

\begin{equation}
\label{l0init2}
    \hat{x} = \operatorname*{argmin}_{x \in \mathbb{R}^N}\|x\|_{p} \ \ subject \ to \ y = Ax .
\end{equation}

The above equation shows an observed signal $y \in \mathbb{R}^M$, an observation matrix $A \in \mathbb{R}^{M\times N}$, and a source signal $x \in \mathbb{R}^N$. Hereafter, the ratio of the number of non-zero entries in $x$ to $N$ is defined as the sparseness $a$, and the ratio of $M$ to $N$ is defined as the compression ratio $\alpha$. Since $l_1$-norm CS is a convex optimisation problem, there are many efficient algorithms for optimisation of $l_1$-norm CS that are widely applied in the real-world problems mentioned above. However, there has been a suggestion that $l_0$-norm CS should outperform $l_1$-norm CS since the $l_1$-norm penalty does not lead to any solution shrinkage \cite{obuchi,kabashima}. In the thermodynamic limit $N$, $M$ $\longrightarrow$ $\infty$ with $\alpha = M/N$ kept fixed, an $l_0$-norm CS’s threshold for $a$, determining whether or not the problem has a solution with no error, is larger than that of $l_1$-norm CS's \cite{obuchi,kabashima}. Nonetheless, the optimisation in $l_0$-norm CS is challenging since it involves combinatorial optimisation.

Numerous attempts have been made to overcome the issue in $l_0$-norm CS optimisations. $l_0$-norm CS can be formulated as a two-fold optimisation \cite{twofold1,twofold2}.

\begin{equation}
\label{l0}
    (\hat{R}, \hat{\sigma}) = \operatorname*{argmin}_{\sigma \in \{0,1\}^{N}}\operatorname*{argmin}_{R\in\mathbb{R}^{N}} \left(\| y - A(\sigma \circ R)\|_{2}^{2}\right) \ \ subject \ to \   \|\sigma\|_{0} \le \Omega .
\end{equation}

Here $R \in \mathbb{R}^N$ and $\sigma \in \left\{{0,1}\right\}^N$ correspond to the source signal and support vector, respectively. 
Especially, each entry in the support vector taking either 0 or 1 represents whether each entry in the source signal is zero or non-zero. The condition $\|\sigma\|_{0} \le \Omega$ is a sparsity-inducing prior for constraining the number of non-zero entries to be $\Omega$. Therefore, the optimisation with respect to $\sigma$ can be regarded as a quadratic-constrained binary optimisation problem to find a ground state of a two-state Potts Hamiltonian. Based on this formulation, simulated annealing (SA) algorithm has been attempted \cite{obuchi}. 
On the other hand, Aonishi \textit{\textit{et al}}., attempted to solve optimisation problems of $l_0$-norm CS with a quantum-classical hybrid approach. $l_0$-norm CS implemented with the hybrid system is given as a regularisation form as follows \cite{Aonishi}.

\begin{equation}
\label{doublel0}
    (R, \sigma) = \operatorname*{argmin}_{\sigma \in \{0,1\}^{N}}\operatorname*{argmin}_{R\in\mathbb{R}^{N}} \left(\frac{1}{2} \| y - A(\sigma \circ R)\|_{2}^{2} + {\lambda} \|\sigma\|_{0}\right) .
\end{equation}

\begin{figure}[!ht]
\centering
     \includegraphics[scale=0.5]{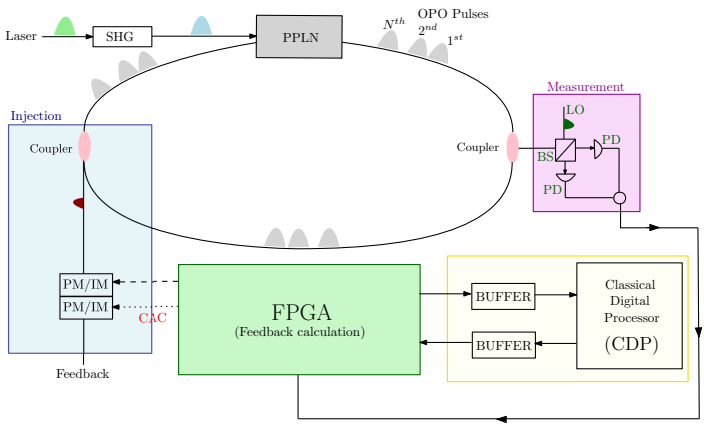}
     \caption{\textbf{CIM-CDP Architecture}}
     \justifying
     {Outline of the system architecture of the 
     the feedback signal including CAC-loop is calculated in FPGA and is fed into the main ring cavity through a coupler. In this hybrid system, CIM optimises the support vector and CDP estimates the source signal in an alternating way.
     Without the CAC-loop, the architecture corresponds to the OL-CIM-CDP while with the CAC-loop, it is the CAC-CIM-CDP. SHG: second harmonic generation, PPLN: periodically poled lithium niobate, BS: beam splitter, PD: photon detector, PM/IM: phase modulator/intensity modulator, LO: local oscillator. 
}
     \label{GACSCIM}
\end{figure}

The element-wise representation of Eq. (\ref{doublel0}) gives the following Hamiltonian.

\begin{equation}
\label{l0Hamiltonian}
    \mathbcal{H} = \sum_{r<r'}^{N}\sum_{k = 1}^{M} A_{r}^{k}A_{r'}^{k}R_{r}R_{r'}\sigma_{r}\sigma_{r'} - \sum_{r=1}^{N}\sum_{k =1}^{M} y^{k}A_{r}^{k}R_{r}\sigma_{r} + {\lambda} \sum_{r = 1}^{N} \sigma_r , 
\end{equation}

where an element $A^k$ in $A$, an element $y^k$ in $y$, an element $R_r$ in $R$ and an element $\sigma_r$ in $\sigma$. Optimisation with respect to $\sigma$ in Eq. (\ref{l0Hamiltonian}) is a quadratic unconstrained binary optimisation (QUBO) problem, which is implementable with a quantum machine such as the coherent Ising machine (CIM) \cite{Aonishi,cimqubo1, cimqubo2, cimqubo3}. In the quantum-classical hybrid approach to conducting $l_0$-regularised CS, $\sigma$ is optimised by the CIM while $R$ is optimised by a Classical Digital Processor (CDP) (see Fig. \ref{GACSCIM}). 

The CIM architecture in the hybrid approach was an open-loop (OL) CIM with the Zeeman term. The hybrid 
approach with the OL-CIM is hereafter referred to as OL-CIM-CDP. Note that the OL means the lack of feedback loop for amplitude control described below. It has been reported that the imbalance in the size of the interaction term and the Zeeman term degrades the system performance \cite{AonishiCDMA}. To balance these terms, for the local field, the measured-amplitudes were binarised. OL-CIM-CDP in this formulation outperformed SA on the regularisation form \cite{Aonishi}.

The close-loop CIM, in which the amplitudes of optical parametric oscillator (OPO) pulses are controlled to a target value, have been proposed to improve the performance of CIM’s ground-state search \cite{kako, Inui2022}. Especially, introducing auxiliary nonlinear dynamics forcefully trying to equalise to a target value results in chaotic behaviour around the target in the CIM which may result in escaping from local minima in the energy landscape. This chaotic method is referred to as chaotic amplitude control (CAC) \cite{AmpLeleu, Leleu1, sam, kako, Inui2022}. Recently, Inui \textit{et al.,} have proposed an approach to efficiently incorporate the Zeeman terms in CAC-CIM by scaling the Zeeman terms with target amplitude to match that of the interaction term \cite{Inui2022}. 

In this paper, following Inui \textit{et al.}’s approach, we modify the CAC-CIM for performing QUBO in $l_0$-regularised CS and attempt to improve the performance of the hybrid CIM-CDP system by replacing the OL-CIM with the CAC-CIM with the Zeeman term (see Fig. \ref{GACSCIM}). The hybrid system proposed here is hereafter referred to as CAC-CIM-CDP. Firstly, to demonstrate the effectiveness of CAC-CIM for performing QUBO in the support estimation, we compare the performance of CAC-CIM to those of OL-CIM and SA. Then, to demonstrate the effectiveness of CAC-CIM-CDP for performing an alternating minimisation, we compare the performance of CAC-CIM-CDP to that of OL-CIM-CDP on artificial random data, as well as magnetic resonance imaging (MRI) data.


\section{Results}
\label{results}

\subsection{Alternating minimisation algorithm}
\label{altmin}

Alternating minimisation procedures on CAC-CIM-CDP and OL-CIM-CDP are summarised in Algorithm \ref{MFBAlgo} and Algorithm \ref{SparseAlgo}, respectively. This type of minimisation suggests the back-and-forth optimisation performed between the CIM and CDP. 
CIM passes the optimisation results to the CDP after optimising the support, as shown in Fig. \ref{GACSCIM}.
The CDP then optimises the signal and sends the resulting signal to the CIM for support optimisation. In Algorithm \ref{MFBAlgo} and Algorithm \ref{SparseAlgo}, indicate the number of iterations of alternating minimisation, the initial values and the integration interval for stochastic differential equations (SDEs) of CIM and so on. The schedules of the pump rate, threshold and target amplitude are given in section \ref{pump}.

\begin{algorithm}
\caption{Alternating minimisation for $l_0$-regularised CS as a QUBO problem on CAC-CIM-CDP. The schedules of the pump rate and threshold are given in Section \ref{pump} }\label{MFBAlgo}
\begin{algorithmic}[1]
\Require $M\times N$ observation matrix: $A,$ $M$-dimensional observation signal: $y$; 
\Ensure $N$-dimensional support vector: $\sigma$, $N$-dimensional signal vector: $r$;
\State Initialise $r = r_{init}$, $\eta = \eta_{init}$, $g^2 =10^{-7}$, ${K} = 1$ and $\tau = 1$
\For{$i = 0$ to $51$}
    \State {Minimise $\mathbcal{H}$ with respect to $\sigma$ by CIM:}
    \State $\sigma$ = CIM-support-estimation$(R, \eta)$ \newline
    \hspace*{2em} Initialise $\tilde{\mu} = 0$, $V = 0.5$ and $e = 1$ for CAC-CIM-CDP (Wigner) and 
    \hspace*{2em} $\tilde{\mu} = 0$, $n = 0$, $m = 0$ and $e = 1$ for CAC-CIM-CDP (Positive-$P$). \newline 
    \hspace*{2em} And we increase the photon’s lifetime 20 times. \newline \hspace*{2em}
    
    \State Minimise $\mathbcal{H}$ with respect to $R$ by CDP using Conjugate Gradient \hspace*{1.2em} Descent or Jacobi method:\newline
    \State Update $\eta$
\EndFor
\end{algorithmic}
\end{algorithm}

\begin{algorithm}
\caption{Alternating minimisation on OL-CIM-CDP. The schedules of the pump rate and threshold are given in Section \ref{pump}}\label{SparseAlgo}
\begin{algorithmic}[1]
\Require $M\times N$ observation matrix: $A,$ $M$-dimensional observation signal: $y$; 
\Ensure $N$-dimensional support vector: $\sigma$, $N$-dimensional signal vector: $r$;
\State Initialise $\tilde{K} = 0.25$, $r = r_{init}$ and $\eta = \eta_{init}$
\For{$i = 0$ to $51$}
    \State {Minimise $\mathbcal{H}$ with respect to $\sigma$ by CIM:}
    \State $\sigma$ = CIM-support-estimation$(R, \eta)$ \newline
    \hspace*{3em} Initialise the in-phase amplitude as $c = 0$, and numerically integrate \hspace*{3em} the W-SDE while increasing the normalised pump rate from 0 to \hspace*{3em} 1.5 for five times the photon’s lifetime when $g^2 = 10^{-7}$.
    \newline \hspace*{2em}
    \State Minimise $\mathbcal{H}$ with respect to $R$ by CDP using Conjugate Gradient \hspace*{1.2em} Descent or Jacobi method:\newline
    \State Update $\eta$
\EndFor
\end{algorithmic}
\end{algorithm}
\bigskip

\subsection{Outline of the CIM models and injection field for QUBO on support estimation}
\label{quboGACS}

On CIM, $l_0$-regularised CS is performed by updating the injection field dictated by the local field, which is determined by the gradient of the QUBO Hamiltonian Eq. (\ref{l0Hamiltonian}) with respect to the spin coordinates. Aonishi \textit{\textit{et al}}., proposed OL-CIM-CDP, which is based on an open-loop injection scheme \cite{Aonishi}. They used the CIM model expressed as the Wigner stochastic differential equation (W-SDE) Eq. (\ref{WSDE0}) and Eq. (\ref{WSDE1}) (in Methods) with the following injection field.  

\begin{equation}
\label{localfieldmain}
    \left(\dfrac{dc_{r}}{dt}\right)_{inj,r} = \left(\abs{h_r} - \eta\right).
\end{equation}

\begin{equation}
\centering
\label{localfieldSparse}
        h_{r} = -{\sum_{r' = 1 (\neq r)}^{N}\sum_{k = 1}^{M}} A_r^k A_{r'}^k R_{r'}H(c_{r'}) + \sum_{k=1}^M A_{r}^k y^{k},
\end{equation}

\begin{figure}[H]
\centering
\includegraphics[scale=0.30]{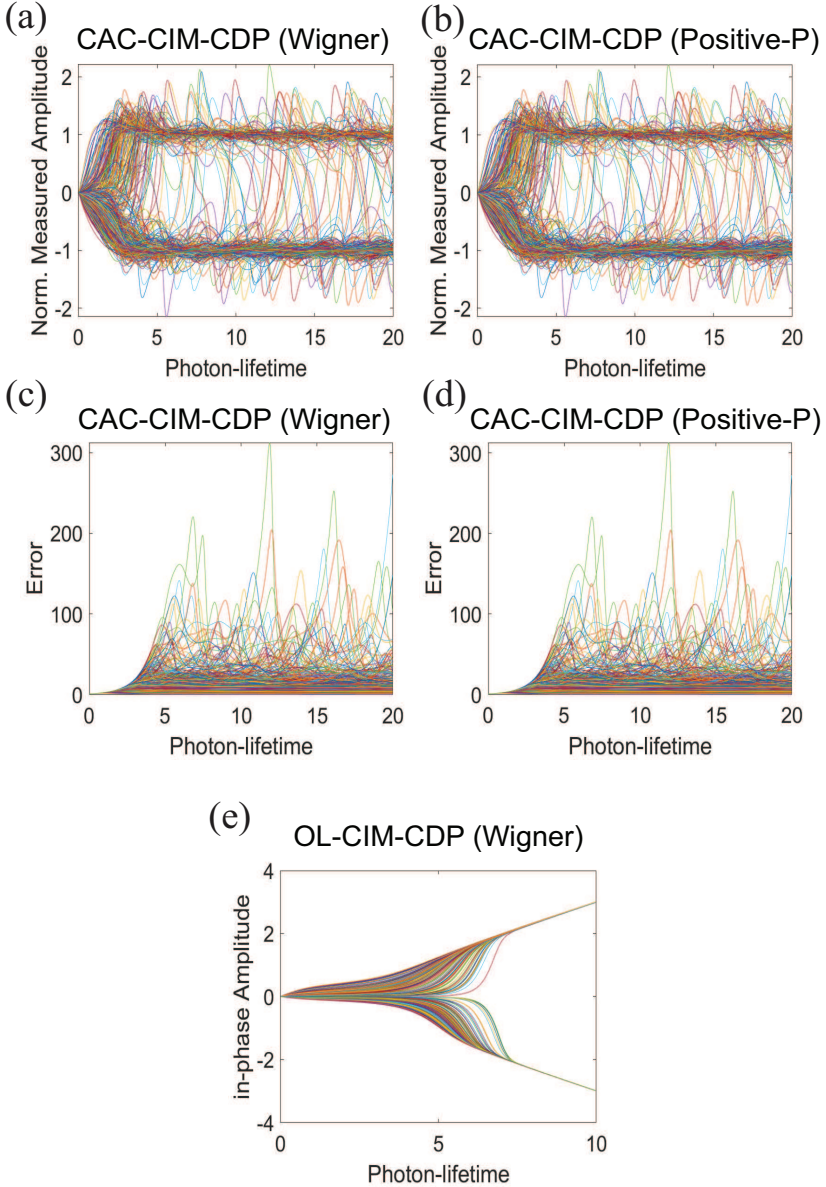}
\caption{\textbf{Amplitude and Error evolution of each CIM model}}
{\begin{justify}\textbf{(a)} and \textbf{(b)} indicates the normalised amplitude $g\tilde{\mu}_r$ evolution of CAC-CIM-CDP (Wigner and Positive-$P$) where $\tau = 1$. With the introduction of CAC to the system, the chaotic behaviour is recognisable in the CAC-CIM-CDP models. \textbf{(c)} and \textbf{(d)} corresponds to the error $e_r$ evolution of CAC-CIM-CDP (Wigner and Positive-$P$). \textbf{(e)} is the in-phase amplitude $c_r$ evolution of OL-CIM-CDP. The system size was set as $N = 2000$ while the compression and the sparseness were 0.6 and 0.2 respectively for all the models.\end{justify}}
\label{MEUgraphs}
\end{figure}

Here, $h_r$ is the local field expressed as Eq. (\ref{localfieldSparse}). $R_r$ is the signal value estimated by the CDP. $c_r$ is the in-phase amplitude of the $r$-th OPO pulse, and $H(c_r)$ is the binarised in-phase amplitude by the Heaviside step function as proposed in the discrete simulated bifurcation \cite{disSimBif}. $\eta$ is the threshold which is related to the $l_0$-regularisation parameter $\lambda$ by $\eta = \sqrt{2\lambda}$ according to the Maxwell rule (see \cite{Aonishi} for a detailed explanation). In the local field Eq. (\ref{localfieldSparse}), the mutual interaction is  $\tilde{J}_{rr'} = -\sum_{k = 1}^M A_r^k A_{r'}^k$ and the Zeeman term is $\sum_{k=1}^M A_{r}^k y^{k}$. Substituting the observation model Eq. (\ref{observationmodelmatrix}) (in Section \ref{observationmodel}) into Eq. (\ref{localfieldSparse}) when $w_{noise} = 0$ (no observation noise), the local field Eq. (\ref{localfieldSparse}) can be expressed as follows.

\begin{equation}
\label{localfield_intro}
    h_r = -{\sum_{r' = 1 (\neq r)}^{N}\sum_{k = 1}^{M}} A_{r}^{k}A_{r'}^{k}R_{r'}H\left(c_{r'}\right) + {\sum_{r' = 1}^{N}\sum_{k = 1}^{M}}A_{r}^{k}A_{r'}^{k}x_{r'} \xi_{r'},
\end{equation}

where $x_r$ is the true signal value, $\xi_r$ is the true support taking 1 or 0. The Zeeman term in the second term of Eq. (\ref{localfield_intro}) can be regarded as the matched filter, in which $A^T A$ is calculated. 
The mutual interaction term in the first term plays a role in removing off-diagonal elements ($r \neq r'$) corresponding to cross-talk noise in the Zeeman term, which are induced by the cross-correlation among the column vectors $A_1,...,A_N$ in $A$. To obliterate the cross-talk noise, the in-phase amplitude $c_r$ needs to be the same as the amplitude of $\xi_r$ if $R_r=x_r$. Hence, $c_r$ is binarised to either 1 or 0. In Fig. \ref{MEUgraphs}e, a typical evolution of $c_r$ in the open-loop-type W-SDE is illustrated. $c_r$ does not keep the same amplitude as that of $\xi_r$ and increases with increasing the pump rate.

In this paper, we propose CAC-CIM-CDP, based on a closed-loop injection scheme with CAC. The idea of CAC for CIM was first introduced by Leleu \textit{et al.,} \cite{AmpLeleu}. It simply states that forcefully trying to equalise the amplitudes of the system to a specific value (in CAC, target amplitude $\tau$) may result in a chaotic behaviour in the system which may result in escaping from local minima in the energy landscape. In this paper, we used two CIM models expressed as W-SDE Eq. (\ref{GACsCIM1}) and (\ref{GACsCIM2}) and Positive-$P$ stochastic differential equation (P-SDE) Eq. (\ref{ppGACsCIM1})-(\ref{ppGACsCIM3}) (in Section \ref{sparsecimmethods}) commonly having the following injection field with CAC feedback.

\begin{equation}
\label{GACSlocalfieldmain}
    \left(\dfrac{d\mu_{r}}{dt}\right)_{inj,r} = je_r\left( R_rh_r - \dfrac{\eta^2}{4}\sqrt{\dfrac{\tau}{g^2}}\right),
\end{equation}
\begin{equation}
\centering
\label{GaussianEC5}
        {\dfrac{d}{dt}e_{r} = -\beta\left(g^2\tilde{\mu}_{r}^2 - \tau\right)e_{r}},
\end{equation}
\begin{equation}
\label{GACsCIM3}
        \tilde{\mu}_{r} = \mu_{r} + \sqrt{\frac{1}{4j}}W_{R,r},
\end{equation}
\begin{equation}
\centering
\label{localfieldGACS}
    h_r = -{\sum_{r' = 1 (\neq r)}^{N}\sum_{k = 1}^{M}} A_{r}^{k}A_{r'}^{k}R_{r'}\dfrac{1}{2}\left(\tilde{\mu}_{r'} + \sqrt{\dfrac{\tau}{g^2}} \right) {+} \sum_{k = 1}^{M} \sqrt{\dfrac{\tau}{g^2}}{A_{r}^{k}y^{k}},
\end{equation}

where $h_r$ is the local field expressed as Eq. (\ref{localfieldGACS}), $e_r$ is the auxiliary variable for the error feedback in the CAC feedback loop, and $\tau$ indicates the target amplitude for the CAC. $R_r$ is the signal value estimated by the CDP, which is the same as that of OL-CIM-CDP. $\eta$ is the threshold given by $\eta = \sqrt{2\lambda}$, which is introduced to keep consistency with OL-CIM-CDP. As described in Section \ref{sparsecimmethods} in Methods, $j$ is the normalised out-coupling rate for optical homodyne measurement, and $g^2$ is the nonlinear saturation parameter of the CIM which determines the abrupt jump of the photon number at the OPO threshold and the amplitude of the quantum noise present in CIM. $\tilde{\mu}_r$ implies the measured-amplitude, and $w (R,r)$ is the independent real Gaussian noise process, which is the same as that in W-SDE (\ref{GACsCIM1}) and P-SDE (\ref{ppGACsCIM1}). In the local field Eq. (\ref{localfieldGACS}), the mutual interaction is  $\tilde{J}_{rr'} = -\sum_{k = 1}^M A_r^k A_{r'}^k$ and the Zeeman term is $h_r^z = \sqrt{{\tau/g^2}}\sum_{k=1}^M A_r^k y^k$. Substituting the observation model Eq. (\ref{observationmodelmatrix}) into Eq. (\ref{localfieldGACS}) when $w_{noise} = 0$ (no observation noise), the local field Eq. (\ref{localfieldGACS}) can be expressed as follows.

\begin{equation}
\label{localfield_introGACS}
    h_r = -{\sum_{r' = 1 (\neq r)}^{N}\sum_{k = 1}^{M}} A_{r}^{k}A_{r'}^{k}R_{r'}\dfrac{1}{2}\left(\tilde{\mu}_{r'} + \sqrt{\dfrac{\tau}{g^2}} \right){ +} {\sum_{r' = 1}^{N}\sum_{k = 1}^{M}} \sqrt{\dfrac{\tau}{g^2}}A_{r}^{k}A_{r'}^{k}x_{r'} \xi_{r'}.
\end{equation}

In Fig. \ref{MEUgraphs}a and \ref{MEUgraphs}b, the typical evolution of normalised measured-amplitude $g\tilde{\mu}_{r}$ are shown. The corresponding error evolution is indicated in Fig. \ref{MEUgraphs}c and \ref{MEUgraphs}d. Due to the CAC feedback loop, as shown in Fig. \ref{MEUgraphs}a and \ref{MEUgraphs}b, if the squared-amplitude of DOPO is smaller than $\tau$, $e_r$ exponentially increases and vice-versa, and the measured-amplitude $\tilde{\mu}_{r'}$ is maintained around $\sqrt{{\tau/g^2}}$. Therefore, because  $1/2(\tilde{\mu}_{r'}+\sqrt{{\tau/g^2}})$ in Eq. (\ref{localfield_introGACS}) can take around $0$ or $\sqrt{{\tau/g^2}}$, the mutual interaction term and the Zeeman term scales are balanced, and crosstalk noise, i.e. off-diagonal elements, is eliminated from the Zeeman term as described in OL-CIM-CDP. Moreover, as shown in Fig. \ref{MEUgraphs}a and \ref{MEUgraphs}b, it is important to note that intermediate solutions are destabilised. By doing so, CAC introduced CIM is able to keep searching for an answer until the maximum run-time has been reached. 
By taking the support vector that is generated by CIM at the end of each trajectory, we are evaluating the solution to estimate the support for the simulations in this paper.

\subsection{Comparison with Simulated Annealing}
\label{SAresults}

Here our purpose is to demonstrate that CAC feedback is effective on CIM by comparing CAC-CIM to OL-CIM and SA.
We follow the Metropolis algorithm for $l_0$-regularised CS stated in \cite{Aonishi}. As same as in \cite{Aonishi}, 1000 samples of the observation matrix and source signal and true support vector are randomly generated according to Section \ref{randomdata} under $N = 500$, $\alpha = a = 0.6$, $w_{noise} = 0$ (no observation noise). With the same observation matrices, source signals, and support vectors in all models, we statistically evaluate how well CAC-CIM estimates support in comparison to OL-CIM and SA when all $R_r$ are fixed to be the source signal $x_r$. To measure the support estimation quality, we used the direction cosine defined as ${\sum_{r=1}^N \xi_r \sigma_r}/{\sqrt{\sum_{r=1}^N \xi_r \sum_{r=1}^N \sigma_r}}$ where $\left(\xi_1, ..., \xi_N\right)$ is the true support vector and $\left(\sigma_1, ..., \sigma_N\right)$ is the estimated one. When the estimation is perfect, the direction cosine is equal to 1. We selected $\eta = 0.05$ corresponding to $l_0$-regularisation parameter $\lambda = \eta^2/2 =0.00125$ as in \cite{Aonishi}.

\begin{figure}[H]
    \begin{center}
    \includegraphics[scale=0.4]{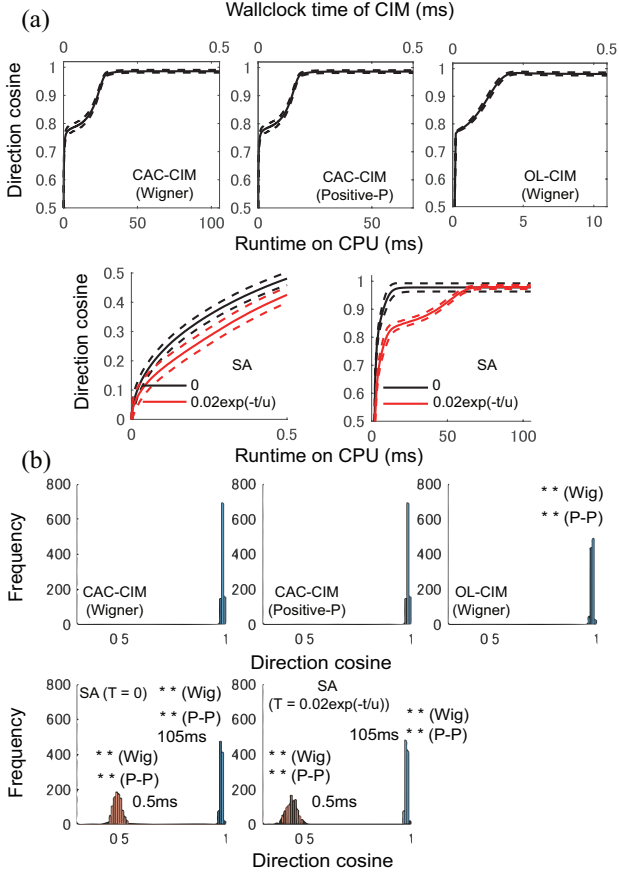}
    \end{center}
    \caption{\textbf{Comparison of CAC-CIM with SA in support vector estimation when SA run-time is set to be the same as that of CAC-CIM (Wigner) model under the same computational environment}}
    \justifying
    {
\textbf{(a)} comparison of the change of direction cosine is mapped for CAC-CIM (Wigner) (upper left), CAC-CIM (Positive-$P$) (upper middle), OL-CIM (upper right) and SA (bottom two). For CAC-CIM the photon's lifetime is increased for 20$\times$ in 105ms of run time. For OL-CIM 5$\times$  photon's lifetime is about 11ms.  In SA 0: constant (zero). $0.02\exp\left(t/u\right)$: exponential cooling scheduling were tested. All graphs show the mean (solid line) and standard deviation (dashed line) of 1000 samples. In our computational environment, the number of Monte Carlo steps for SA with a run time of 105ms is about 46000.
\textbf{(b)} Histogram of 1000 final states of the direction cosine is shown (CAC-CIM (Wigner) (upper left), CAC-CIM (Positive-$P$) (upper middle), OL-CAC (upper right) and SA (bottom two)). ** (Wig) and **(P-P) in the graphs means that cumulative histograms of these final states are significantly higher than Wigner and Positive-$P$ models (P-value $<$ 0.0001 on two-sample one-sided Kolmogorov-Smirnov test) and thus the final states of Wigner and Positive-$P$ models are biased towards 1 compared to those final states. $N=500$, $\alpha=a=0.6$. In all CIM, $g^2=10^{-7}$. }
    \label{SAgraphs}
\end{figure}

First, we evaluate the temporal profiles of the optimisation processes for the support estimation in CAC-CIM (Wigner), CAC-CIM (Positive-$P$), OL-CIM and SA. The upper three graphs (from left to right, CAC-CIM (Wigner), CAC-CIM (Positive-$P$) and OL-CIM respectively) in Fig. \ref{SAgraphs}a show the change in the direction cosine of the three CIM models depending on the runtime on the CPU and the {wall-clock time} of physical CIM. For CAC-CIM (Wigner), and CAC-CIM (Positive-$P$) models, $20\times$ photon’s lifetimes of integral interval (with 1000 time-steps) for the SDEs are about 105ms and 68ms of run-time respectively, and for OL-CIM, $5\times$ photon’s lifetime of integral interval (with 50 time-steps) for the SDE is about 11ms. The physical CIM's wall-clock time for this optimisation is roughly estimated to be around 0.5ms, which can be estimated from the round-trip time of $N=500$ and the time-steps-to-solution for the Sherrington-Kirkpatrick problem with $N = 500$ \cite{sam}. The direction cosine of these CIM models converged to about 1 by these run-times. The lower two graphs in Fig. \ref{SAgraphs}a show the change in the direction cosine of SA depending on the runtime on CPU under constant temperature at $T=0$ and exponential cooling scheduling from $T=0.02$ to $0.00002$. We adjusted the Monte-Carlo steps of SA (bottom two graphs of Fig. \ref{SAgraphs}a) to accompany the wall-clock time of physical CIM (0.5ms) and the run-time of CAC-CIM (Wigner) (105ms). In our computational environment, the number of Monte Carlo steps for SA with runtimes of 0.5ms and 105ms is about 230 and 46000 steps, respectively. In SA, the direction cosine converged to about 1 by 105ms, while that did not by 0.5ms.

Next, we compare the histogram of the final states of direction cosines in CAC-CIM (Wigner), CAC-CIM (Positive-$P$), OL-CIM and SA. The upper three graphs in Fig. \ref{SAgraphs}b indicate the histogram of the three CIM models (CAC-CIM (Wigner), CAC-CIM (Positive-$P$), OL-CIM, respectively), while the lower two graphs in Fig. \ref{SAgraphs}b show the histograms of SA at run-times of 0.5ms and 105ms under zero temperature and exponential cooling schedules respectively. Comparing these graphs, the proportion of the direction cosines of CAC-CIM (Wigner) and CAC-CIM (Positive-$P$) close to 1 is higher than those of OL-CIM and SA. The two-sample one-sided Kolmogorov-Smirnov test suggests that the histograms of the final direction cosines of CAC-CIM (Wigner) and CAC-CIM (Positive-$P$) are significantly biased toward 1 compared with all of those of OL-CIM and SA (P-value $<$ 0.0001).

The above results thus demonstrate that CAC-CIM outperformed OL-CIM on support vector estimation and outperformed SA within the same run-time.

\subsection{Comparison with ground state predicted with statistical mechanics on alternating minimisation}
\label{noisy}


We compare CAC-CIM-CDP's capability to find the ground state with that of OL-CIM-CDP. In our previous study, we derived the macroscopic parameter equation (Eq. (26)-(28) in \cite{Aonishi}) using a non-equilibrium statistical mechanics method to show the performance limit of OL-CIM-CDP. In the limit of the saturation parameter $g^2 \rightarrow 0$, the macroscopic parameter equation derived in the previous study is consistent with that for a two-state Potts spin system defined by the QUBO Hamiltonian Eq. (\ref{l0Hamiltonian}). Therefore, the macroscopic parameter equation in this limit can predict the ground state of the Hamiltonian. Through a comparison of solutions of CAC-CIM-CDP and OL-CIM-CDP with a solution of the macroscopic parameter equation in the limit of $g \rightarrow 0$, we demonstrate the efficacy of CAC feedback on the alternating minimisation for optimising the Hamiltonian.

\begin{figure}[H]
\centering
\includegraphics[scale=0.29]{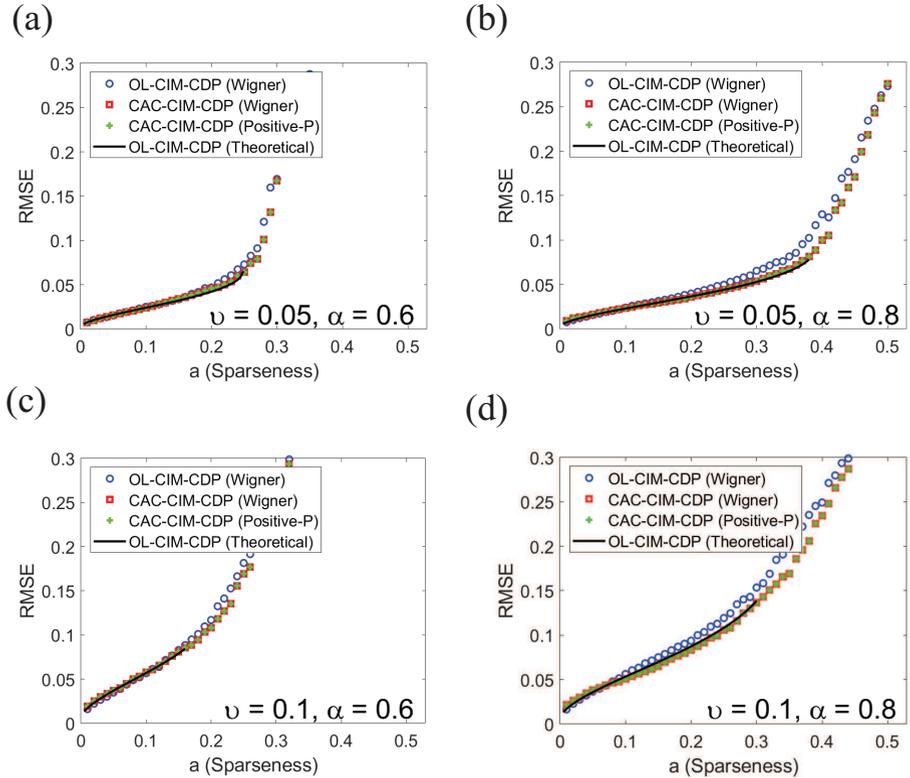}
\caption{\textbf{Comparison of average RMSE of CAC-CIM-CDP models to the theoretical limit of OL-CIM-CDP when observation noise is present}}
\centering
\begin{justify}
{\textbf{(a)} and \textbf{(b)} indicates the average performance for $N = 2000$ system where $\alpha = 0.6$ and $\alpha = 0.8$ respectively for $\nu = 0.05$. \textbf{(c)} and \textbf{(d)} states the average performance for $\nu = 0.1$. 
For all graphs $\eta_{init} = 0.8$ and $\eta_{init} = 0.6$ was used for CAC-CIM-CDP models and OL-CIM-CDP respectively. \textbf{(a)} and \textbf{(b)} $\eta_{end}$ was set to 0.18. \textbf{(c)} and \textbf{(d)} $\eta_{end}$ was set to 0.35.}
\end{justify}
\label{ppRMSETarggraphs}
\end{figure}

The precondition for applying statistical mechanics is that the values of all entries in the observation model Eq. (\ref{observationmodelmatrix}), which is the premise of Eq. (\ref{doublel0}) and Eq. (\ref{l0Hamiltonian}), are randomly determined as described in Section \ref{randomdata}. To compare solutions of the models with the ground state predicted with statistical mechanics, 10 samples of the observation matrix and source signal and true support vector are randomly generated according to Section \ref{randomdata} under $N = 2000$ and various values of $a, \alpha$ and $\nu$. Here $\nu$ indicates the standard deviation of the observation noise ($w_{noise}$). Then, we execute Algorithms \ref{MFBAlgo} and \ref{SparseAlgo} for the alternating minimisation in CAC-CIM-CDPs (Wigner and Positive-$P$) and OL-CIM-CDP sharing the same samples of observation matrices, source signals and support vectors. Here for Fig. \ref{ppRMSETarggraphs}, $\eta_{init} = 0.6$ and $\eta_{init} = 0.8$ was used for CAC-CIM-CDP models and OL-CIM-CDP respectively. $\eta_{end}$ was set to 0.18 in Fig. \ref{ppRMSETarggraphs}a and Fig. \ref{ppRMSETarggraphs}b while in  Fig. \ref{ppRMSETarggraphs}c and  Fig. \ref{ppRMSETarggraphs}d $\eta_{end}$ was set to 0.35. 

The marks in Fig. \ref{ppRMSETarggraphs} show the averaged root-mean-square-error (RMSE) calculated as $\sqrt{1/N \sum_{r=1}^N \left(R_r\sigma_r - x_r\xi_r\right)^2}$ of sampled solutions obtained from OL-CIM-CDP, Wigner and Positive-$P$ of CAC-CIM-CDPs. Here $\sigma_r$ is calculated as stated in Eq. (\ref{heaviside}). The black solid lines in Fig. \ref{ppRMSETarggraphs} indicate RMSE at the ground state corresponding to successful signal retrieval, which is predicted with statistical mechanics. RMSEs of Wigner and Positive-$P$ CAC-CIM-CDPs tend to keep a better consistency with that of the ground state compared to OL-CIM-CDP for various values of $a, \alpha$ and $v$. Especially as shown in Figs. \ref{ppRMSETarggraphs}b and \ref{ppRMSETarggraphs}d, RMSE of OL-CIM-CDP tend to deviate gradually from that of the ground state as increasing $a$, while both Wigner and Positive-$P$ CAC-CIM-CDPs keep up a better consistency with the theoretical prediction.  

\subsection{Application to Sparse MRI}
\label{mriSimulations}

We evaluate the performance of CAC-CIM-CDP, OL-CIM-CDP and LASSO \cite{Tibshirani} on MRI data. 

In the following numerical experiment, we used two different-sized sparse images ($64\times64$ and $128\times128$ pixels) spanned by a Haar basis function. Detailed explanations of the two images we used as the source images are given in Section \ref{MRIdata} in Methods. In accordance with our previous work \cite{Aonishi}, we sought to reconstruct the two images from the undersampled $k$-space data and by solving the optimisation problem defined in Eq. (\ref{MRIl0init1}) (see Section \ref{MRIdata}). To realise the optimisation problem in Eq. (\ref{MRIl0init1}) on CIM, the Haar wavelet transform coefficients are estimated with the mutual interaction term and the Zeeman term constructed according to Eq. (\ref{MRIzeeman}) and (\ref{MRIJMatrix1}) in Section \ref{MRIdata}. The compression rate of the $k$-space data from the $64\times64$ and $128\times128$ images is 0.4 and 0.3 respectively. And the sparseness of the images is 0.212 and 0.178 respectively. 
As the solver for CDP, we used the Conjugate Gradient Descent method (further details on CDP optimisation refer to Section \ref{optCDP}). 

\begin{figure}[H]
\centering
\includegraphics[scale=0.55]{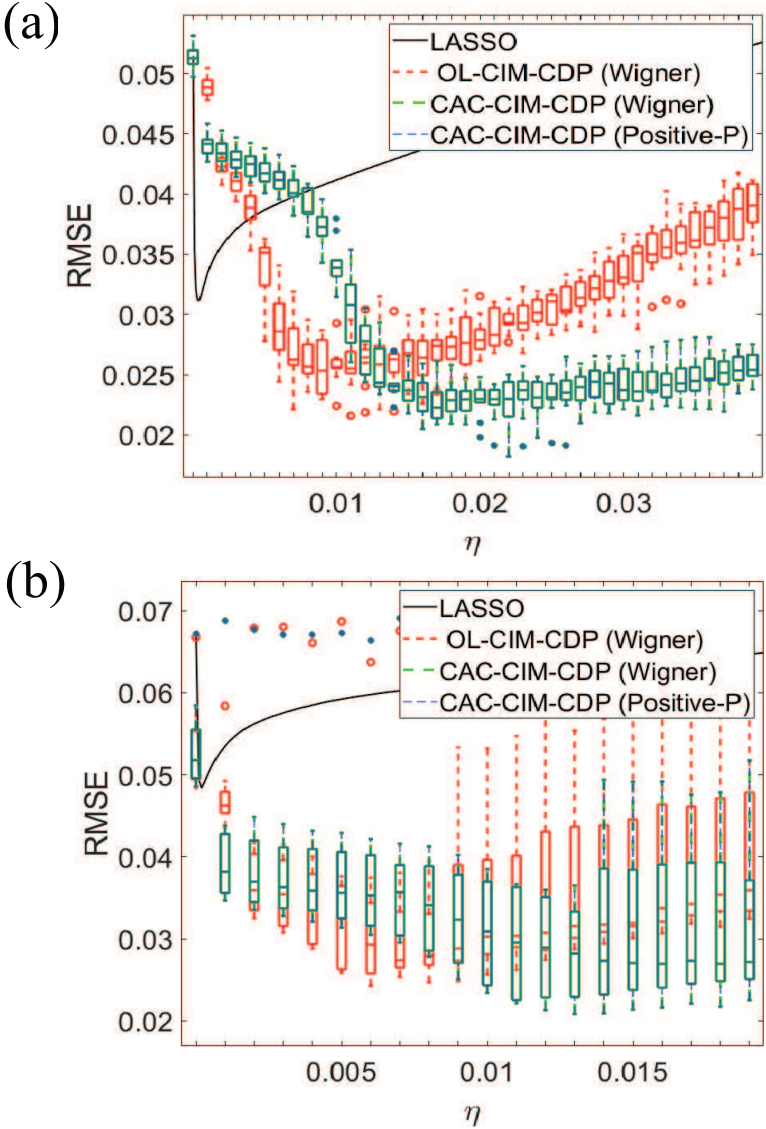}
\caption{\textbf{Average performance of the models when $l_0$-regularisation parameter varies for different image sizes}}
\begin{justify}
\textbf{(a)} Performance on $64\times64$ and \textbf{(b)} Performance on $128\times128$. The black line indicates the performance on LASSO while the red boxes correspond to OL-CIM-CDP. Green and blue boxes indicate the performance on CAC-CIM-CDP Wigner and Positive-$P$ respectively. For different threshold values, 
the graphs illustrate the maximum, minimum, 25-th percentile (bottom edge), 75-th percentile (top edge), and median (central horizontal line) of RMSEs for each model with box plots. {The markers indicate the outliers.}
The compression and sparseness for \textbf{(a)} were 0.4 and 0.212 respectively while for \textbf{(b)} were 0.3 and 0.178.
\end{justify}
\label{mriAllgraphs}
\end{figure}
\begin{figure}[H]
\centering
\includegraphics[scale=0.4]{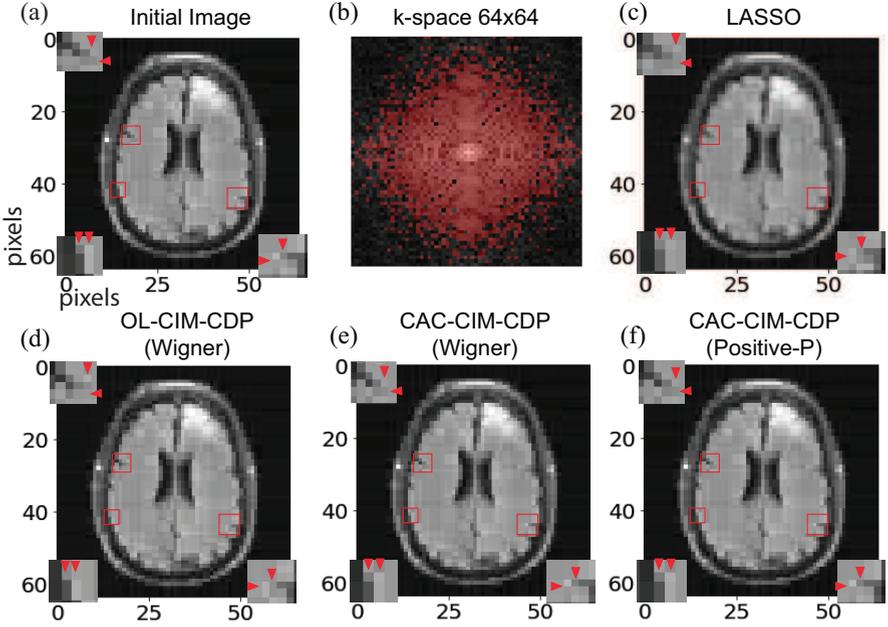}
\caption{\textbf{Reconstructed Images for $64\times64$}}
\centering
\begin{justify}
\textbf{(a)} Resized $64\times 64$ initial image. The compression and sparseness were 0.4 and 0.212 respectively. \textbf{(b)} Undersampled k-space data (random red points). \textbf{(c)}, \textbf{(d)}, \textbf{(e)}, and \textbf{(f)} correspond to the reconstructions obtained from LASSO, OL-CIM-CDP, CAC-CIM-CDP (Wigner), and CAC-CIM-CDP (Positive-$P$) with RMSE values 0.0292, 0.0216, 0.0182 and 0.0182 respectively. The enlarged image portions indicate the pixel-wise differences between the reconstructions. For \textbf{(d)} 31 alternating minimisation processes were performed. For \textbf{(e)} and \textbf{(f)} 11 alternating minimisations were performed. And for \textbf{(c)}, \textbf{(d)}, \textbf{(e)}, and \textbf{(f)} $\eta_{init} = \eta_{end}$ was 0.0003, 0.011, 0.022, and 0.022 respectively. 
\end{justify}
\label{mriImagegraphs64}
\end{figure}
\begin{figure}[H]
\centering
\includegraphics[scale=0.29]{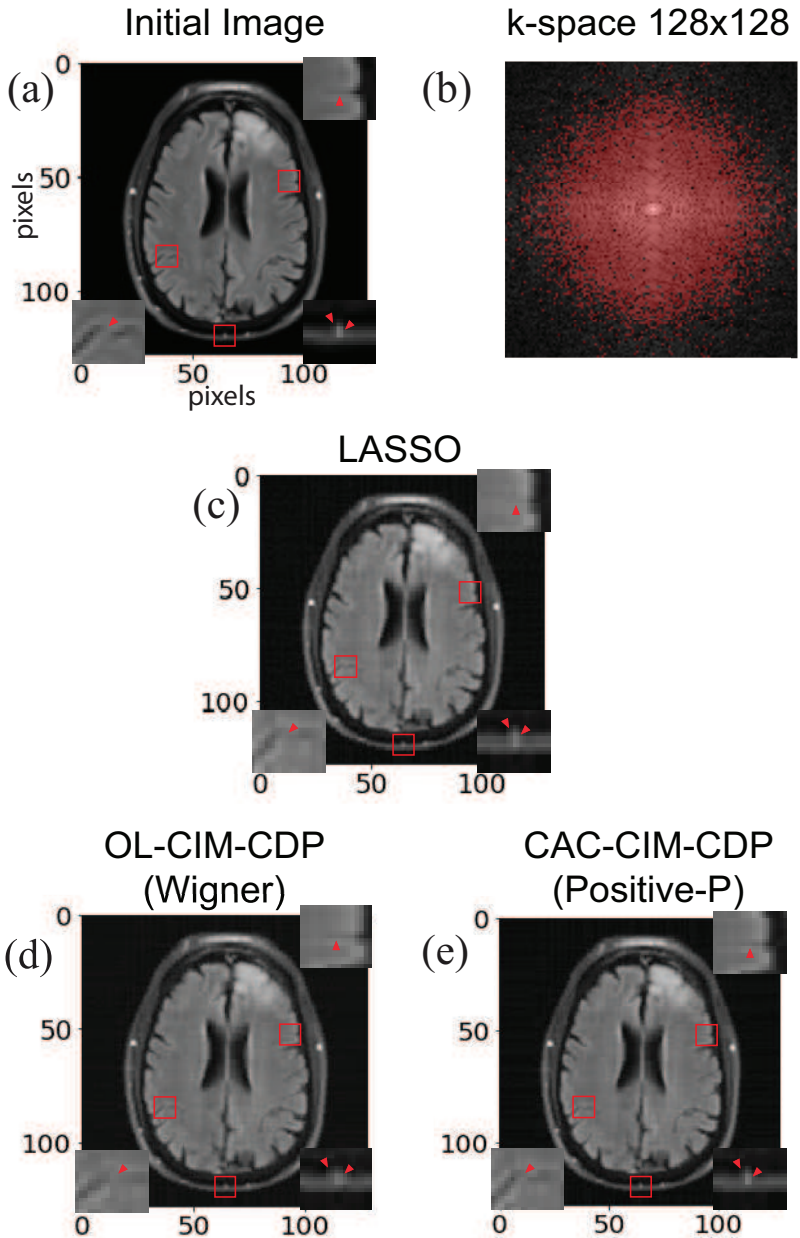}
\caption{\textbf{Reconstructed Images for $128\times128$}}
\centering
\begin{justify}
\textbf{(a)} Resized $128\times 128$ initial image. The compression and sparseness were 0.3 and 0.178 respectively
\textbf{(a)} Red dots indicate the sampled k-space from the \textbf{(a)}'s k-space. 
\textbf{(c)}, \textbf{(d)}, and \textbf{(e)} correspond to the reconstructions obtained from LASSO, OL-CIM-CDP, and CAC-CIM-CDP (Positive-$P$) with RMSE values 0.0276, 0.0243, and 0.0209 respectively.
The enlarged image portions indicate the pixel-wise differences between the reconstructions. For \textbf{(d)} and \textbf{(e)}, 31 and 11 alternating minimisation processes were performed respectively. And for \textbf{(c)}, \textbf{(d)}, and \textbf{(e)} $\eta_{init} = \eta_{end}$ was 0.0001, 0.006, and 0.013 respectively. 
\end{justify}
\label{mriImagegraphs128}
\end{figure}

In Fig. \ref{mriAllgraphs}a and Fig. \ref{mriAllgraphs}b, for 10 simulations the average RMSE value is indicated for each threshold $\eta$ for $64\times64$ and $128\times128$ images respectively. As for the minimum RMSE in the $64\times 64$ case, LASSO (black line), OL-CIM-CDP (red), CAC-CIM-CDP (Wigner) (green) and CAC-CIM-CDP (Positive-$P$)'s (blue) can be stated as, 
0.0292, 0.0216, 0.0182 and 0.0182 respectively (for the corresponding reconstructions see Fig. \ref{mriImagegraphs64}).
In the $128\times128$ case, the minimum RMSE is 
0.0276, 0.0242, 0.0209 and 0.0209 respectively (for the corresponding reconstructions see Fig. \ref{mriImagegraphs128}). Comparing the RMSE values acquired it is clear that CAC-CIM-CDP models have a better average performance compared to the other approaches in both image sizes. And even after reaching the optimal reconstruction for the given parameters, CAC-CIM-CDP tends to keep up a minimal error rate compared to LASSO and OL-CIM-CDP. This indicates that the effective range of CAC-CIM-CDP is much wider than OL-CIM-CDP. 
In both image sizes, the Wigner and Positive-$P$ variations of CAC-CIM-CDP produce identical RMSE results.


In Fig. \ref{mriImagegraphs64} and \ref{mriImagegraphs128} the minimal RMSE constructions are shown for LASSO, OL-CIM-CDP, CAC-CIM-CDP (Wigner) and CAC-CIM-CDP (Positive-$P$). In Fig. \ref{mriImagegraphs128}, only CAC-CIM-CDP (Positive-$P$)'s reconstruction is shown because it is clear that both CAC-CIM-CDP (Wigner) and CAC-CIM-CDP (Positive-$P$)'s performance is identical. In the $64\times64$ image reconstruction when RMSE values are compared, CAC-CIM-CDP models have better reconstruction accuracy. The enlarged portions indicate the difference in pixel identification of each model compared to the initial resized image. Considering both simulations it is clear that even though the system size increases, proposing models have the upper hand in performing an accurate reconstruction compared to other models.

\section{Discussion}

In this paper, we have proposed an improved CIM approach to solve $l_0$-regularised compressed sensing problems. The proposed algorithm has shown that it can outperform the previously proposed algorithm accuracy-wise in all the simulations performed. With the OL-CIM algorithm, the CIM model in use was lacking the CAC feedback for chaotically exploring solutions. 
Therefore, CAC-CIM has been able to provide convergence to a better solution than OL-CIM. One factor to emphasise here is that CAC does not guarantee convergence to the ground state. Even the ground state is reached, due to the forceful equalisation to $\tau$ may prevent from stopping there. Even though this is the case in this paper, CAC has been shown to be effective especially when the problem instances are relatively harder in both artificial random data and MRI data. 

\subsection{Effect of system size on performance}

The introduction of CAC has previously been shown to have better performance with small-scale frustrated Ising problem instances \cite{Inui2022}. In this manuscript, we have demonstrated the applicability of CAC for real-world combinatorial optimisation problems (in this case Compressed sensing) where the problem instances with a Zeeman term are mapped to a QUBO formulation that is large-scale. The simulations with random artificial data on various system sizes are illustrated in Supplementary note 1.
Even though the performance increase is present, in very large system sizes such as in $128\times 128$, it is clear that the RMSE gap between CAC-CIM-CDP and OL-CIM-CDP is smaller compared to $64\times 64$. This poses the question that whether there is a system-size threshold for CAC-CIM-CDP in the very-large-scale regime. Considering the MRI-based simulations require 4096 and 16384 DOPO pulses to operate (compared to 16 DOPOs in theoretical simulations in \cite{Inui2022}), the system size of CAC’s applicability is largely improved. Yet the system-size-wise dependency is yet to be explored. 



\subsection{Advantages of CAC-CIM architecture}

With the use of CDP, the problem which involves quadratic optimisation has been solved in this hybrid system. As shown in the schematic illustration of the CAC-CIM-CDP in Fig. \ref{GACSCIM}, proposing approach performs an alternating minimisation between the CIM and CDP. It is clear considering the results stated in Section \ref{mriSimulations} that CAC-CIM-CDP has outperformed OL-CIM-CDP and the generally used approach LASSO which is an $l_1$-regularised method for solving compressed sensing problems. 
It is interesting to see that advancements in CIM architecture can offer better results in real-world problem instances.


\subsection{CAC-CIM-CDP (Wigner) Vs. CAC-CIM-CDP (Positive-$P$)}

Even though this paper introduces two variants (Wigner and Positive-$P$) of CAC-CIM-CDP, the performances have been almost identical between the models. However, we encountered a deviation when the problem instances become harder i.e. sparseness/compression ratio becomes higher when $w_{noise} = 0$. The results are presented in Supplementary note 2.
As the models approach a threshold point for optimal reconstruction (a critical sparseness/compression ratio), beyond that the producing RMSE values are somewhat different between the models. Performance-wise it is hard to state that one model is better than the other. Because the significance of Wigner and Positive-$P$ lies in the density matrix approximation and how it behaves with a large quantum noise presence. We discuss this in Supplementary note 2.

\subsection{Future improvements to the CAC-CIM-CDP}

\subsubsection{Simultaneous minimisation}

One of the major bottlenecks the proposed model (CAC-CIM-CDP) has is the alternating minimisation process between the CIM and CDP. This is a time-consuming operation. As a future direction to this model, we plan to improvise the CIM system to accommodate quadratic optimisation problems and perform simultaneous minimisation using only the CIM to solve compressed sensing problems. We believe that the use of "CIM-only" will have a positive effect on accuracy as well. 

\subsubsection{CAC-CIM-CDP with large quantum noise}

While this manuscript solely focuses on combining CAC with CIM for solving CS problems more accurately, the considered quantum noise present in the CIM is very low ($g^2 = 10^{-7}$). This opens up a problem of whether CAC-CIM-CDP can keep up the performance with a large quantum noise presence. For small-scale frustrated Ising hamiltonians, this has been previously explored in \cite{Inui2022} ($N = 16$) where it has shown a decrease in success probability for larger $g^2$ terms. This result is consistent with CAC-CIM-CDP as well as shown in Supplementary note Fig. \ref{mri64saturationgraphs} for MRI simulations. Recent advances in CIM research have led to the introduction of a method known as Negative Parametric Gain (NPG), which accommodates higher quantum noise and at the same time as maintaining a higher probability of success \cite{npg}. This method considers a negative starting pump rate with large injection field feedback. NPG has shown promising results in the theoretical simulations \cite{npg}. We are planning to improve the endurance of the CAC-CIM-CDP with NPG for a larger quantum noise presence. 

\subsubsection{CAC-CIM-CDP with the mean-field CIM model}

As it is obvious from the perspective of numerical simulations, CAC-CIM-CDP SDEs are computationally costly to simulate. Even though the shown results are acquired using a GPU implementation of the SDEs, as a digital simulator, field-programmable gate arrays (FPGAs) are more suitable (less energy cost, faster processing etc). As a future direction, we plan on implementing the mean-field CIM SDEs \cite{sam, AmpLeleu} with CAC on an FPGA to perform compressed sensing simulations. Due to the fact that CAC-CIM-CDP has relatively low noise present in the system, we believe that the mean-field SDEs will have approximately the same or better results but with faster simulation times. This is mainly due to the simplicity and the negligence of the noise terms in the mean-field CIM SDEs. 

\section{Methods}\label{methods}

\subsection{Stochastic Differential equation in OL-CIM-CDP and CAC-CIM-CDP}
\label{sparsecimmethods}

\subsubsection{Wigner-type}


The CIM model based on the Wigner formulation was introduced in \cite{nonoiseCIM,Inui}. The $c$-number Heisenberg Langevin equation \cite{nonoiseCIM} was used to overcome the higher computational cost of simulating the direct density matrix formulation of CIM and it has been found to be equivalent to the truncated Wigner SDEs. The density operator master equation expanded by the Wigner function results in the Kramers-Moyal series including third-order terms. In order to derive the Langevin equation, we neglect third-order terms \cite{Inui2022}. Then, we can formulate the following Wigner SDEs used for OL-CIM-CDP.

\begin{equation}
\label{WSDE0}
\begin{multlined}
        \frac{d}{dt}c_r = \left[-1 + p - {\left(c_r^2 + s_r^2\right)} \right]c_r + \widetilde{K}\left(\dfrac{dc_{r}}{dt}\right)_{inj,r} +\\ {g}\sqrt{\left(c_r^2 + s_r^2\right) + \frac{1}{2}} W_{r,1},
\end{multlined}
\end{equation}
\begin{equation}
\label{WSDE1}
\begin{multlined}
        \frac{d}{dt}s_r = \left[-1 - p - {\left(c_r^2 + s_r^2\right)}\right]s_r + {g}\sqrt{\left(c_r^2 + s_r^2\right) + \frac{1}{2}} W_{r,2} .
\end{multlined}
\end{equation}

Here, in-phase and quadrature-phase normalised amplitudes are represented as $c_r$ and $s_r$ respectively. $p$ is the normalised pump rate. If $p$ is above the oscillation threshold $(p > 1)$, each of the OPO pulses is either in the $0$-phase state or $\pi$-phase state. The last terms of the upper and lower equations express the vacuum fluctuations injected from external reservoirs and the pump fluctuations coupled to the OPO system via gain saturation \cite{Aonishi}. $W_{r,1}$ and $W_{r,2}$ are independent real Gaussian noise processes satisfying $\langle W_{r,k} (t)\rangle =0$ and $\langle W_{r,k}(t) W_{r',l} (t')\rangle = \delta_{rr'} \delta_{kl} \delta(t-t')$. ${g}$ indicates the saturation parameter. $(dc_r/dt)_{inj,r}$ is the optical injection field, which only considers the in-phase amplitudes for the calculations. The injection field is defined in Eq. (\ref{localfieldmain}) and Eq. (\ref{localfieldSparse}). $\Tilde{K}$ indicates the normalised feedback strength. 

Focusing on the behaviour of the OPO pulses only in the in-phase direction, the Wigner-type SDE, which is used for CAC-CIM-CDP, can be stated as,

\begin{equation}
\label{GACsCIM1}
\begin{multlined}
        \dfrac{d}{dt}\mu_{r} = - \left(1 -p + j\right)\mu_{r} - g^2\mu_{r}^3 + \sqrt{j}\left(V_{r} - \frac{1}{2}\right)W_{R,r} + {{K}}\left(\frac{d\mu_{r}}{dt}\right)_{inj,r},
\end{multlined}
\end{equation}
\begin{equation}
\begin{multlined}
\label{GACsCIM2}
        \dfrac{d}{dt}V_{r} = -2 \left(1 -p + j\right)V_{r} - 6g^2\mu_{r}^2V_{r} + 1 + j + 2g^2\mu_{r}^2 - 2j\left(V_{r} -\frac{1}{2}\right)^2 .
\end{multlined}
\end{equation}

Here $\mu_r$ and $V_r$  are the mean-amplitudes and the variance of the $r$-th DOPO pulse. $(d\mu_r/dt)_{inj,r}$ is the optical injection field defined in Eq. (\ref{GACSlocalfieldmain})-(\ref{localfieldGACS}). $W_{R,r}$ is independent real Gaussian noise processes satisfying $\langle W_{R,r} (t)\rangle =0$ and $\langle W_{R,r} (t)W_{R,r'} (t')\rangle =\delta_{rr'}\delta(t-t')$. $g$, $p$, $j$ and ${K}$ indicate the saturation parameter, pump rate, the normalised out-coupling rate for optical homodyne measurement and the feedback {strength}, respectively. 

\subsubsection{Positive-$P$-type}

Positive-$P$ (P-P) representation \cite{drumOps} is a generalised form of Glauber–Sudarshan $P$ representation. When the density operator master equations are expanded using the P-P distribution function, the resulting Kramers-Moyal series only consists of first and second-order terms. Due to this factor, there is no truncation needed to derive the Langevin equation. Because of this one can argue that P-P SDEs might be a better candidate for density operator approximations. The effectiveness of P-P SDEs has been demonstrated on CIMs with higher quantum noise presence \cite{Inui2022}. We can formulate the P-P-type SDEs we used for CAC-CIM-CDP.

\begin{equation}
\label{ppGACsCIM1}
\begin{multlined}
        \dfrac{d}{dt}\mu_{r} = - \left(1 -p + j\right)\mu_{r} - {g^2\mu_{r}\left(\mu_{r}^{2} + 2n_r + m_r\right)} + \sqrt{j}\left(m_r + n_r\right)W_{R,r}\\ + {{K}}\left(\frac{d\mu_{r}}{dt}\right)_{inj,r},
\end{multlined}
\end{equation}
\begin{equation}
\label{ppGACsCIM2}
\begin{multlined}
        {\dfrac{d}{dt}n_{r} = -2 \left(1 + j\right)n_r + 2pm_r - 2g^{2}\mu_{r}^2\left(2n_r + m_r\right)} {- j\left(m_r + n_r\right)^{2}},
\end{multlined}
\end{equation}
\begin{equation}
\label{ppGACsCIM3}
\begin{multlined}
        {\dfrac{d}{dt}m_{r} = -2 \left(1 + j\right)m_r + 2p n_r - 2g^{2}\mu_{r}^{2}\left(2m_r + n_r\right) + } p\\ -  g^{2}\left(\mu_{r}^{2} + m_r\right) - j\left(m_r + n_r\right)^{2} .
\end{multlined}
\end{equation}

Here $\mu_r$ corresponds to the mean-amplitude, $m_r$ and $n_r$ represent variances of quantum fluctuations of the $r$-th DOPO pulse. $(d\mu_r/dt)_{inj,r}$ is the optical injection field defined in Eq. (\ref{GACSlocalfieldmain})-(\ref{localfieldGACS}). $W_{R,r}$ is independent real Gaussian noise processes satisfying $\langle W_{R,r} (t)\rangle =0$ and $\langle W_{R,r} (t)W_{R,r'} (t')\rangle =\delta_{rr'}\delta(t-t')$. $g$,$ p$, $j$ and ${K}$ are the same as those for the Wigner model.  

\subsection{Optimisation in CDP}
\label{optCDP}

The CDP performs the optimisation of the Hamiltonian (Eq. \ref{l0Hamiltonian}) with respect to $R_r$ for a support vector $\sigma$ given by CIM. $\sigma$ is obtained by binarising the measured-amplitude ($\Tilde{\mu}_r$) defined in Eq. (\ref{GACsCIM3}) (CAC-CIM-CDP) or in-phase amplitude $c_r$ (OL-CIM-CDP) with the Heaviside function stated as, 

\begin{equation}
\label{heaviside}
    \sigma_{r} = Heaviside\left(x_r\right)= 
\begin{cases}
    1,&  \ \left(x_r > 0\right)\\
    0,&   \ \left(x_r \leq 0\right) .
\end{cases}
\end{equation}

The CDP solve the following system of equations, which is satisfied the stationary point that minimises $\mathbb{H}$ with respect to $r$. 

\begin{equation}
\label{localfieldCDP1}
      R_{r}\sum_{k = 1}^{M} \left(A_{r}^{k}\right)^2 = \sigma_{r}\mathbb{H}_{r},
\end{equation}
\begin{equation}
\label{localfieldCDP2}
    \mathbb{H}_{r} = -\sum_{r' = 1 (\neq r)}^{N}\sum_{k = 1}^{M} A_{r}^{k}A_{r'}^{k}R_{r'}\sigma_{r'} + \sum_{k =1}^{M} A_{r}^{k}y^{k} .
\end{equation}

Here, $\mathbb{H}_r$ in Eq. (\ref{localfieldCDP2}) is the local field of the CDP, which is the same as Eq. (\ref{l0Hamiltonian}) and (\ref{localfieldGACS}). For the simulations, we used the Jacobi method or Conjugate Gradient Descent (CGD) method as the CDP optimiser. During the optimisation in the CDP, all $\sigma_r$ are fixed.

\subsection{Schedule of pump rate, threshold and target amplitude for optimisation in CIM}
\label{pump}

A rough parameter search was used to determine the schedules for each of the following parameters in the experiments.
The pump rate $p$ for both Wigner and P-P type CAC-CIM-CDPs was scheduled depending on the time $t$ as follows.

\begin{equation}
\label{pumprateGACS}
        p = (p_{thr} - d) + \frac{2d}{1+e^{-\left(\dfrac{t-4}{2}\right)}} .
\end{equation}

Here, $p_{thr} = 1$ for all simulations of both Wigner and P-P type CAC-CIM-CDPs. For artificial random data and MRI data simulations, $d$ was set at 0.6 and 0.4 respectively. 

In accordance with \cite{Aonishi}, the pump rate $p$ for OL-CIM-CDP was scheduled depending on the time $t$ as follows.

\begin{equation}
\label{pumprateMFB}
        p =  1.5 \times \left(\dfrac{t}{5}\right)^2 .
\end{equation}

The pump rate becomes equal to $1.5$ when $t=5$. We used this pump rate schedule for all simulations of OL-CIM-CDP.
In both CAC-CIM-CDP and OL-CIM-CDP, the threshold $\eta$ was scheduled depending on the alternating iteration time $i$ as follows.

\begin{equation}
\label{etaSchedulling}
        \eta_{i} = \max \left[\eta_{init} \left(1 - \dfrac{i}{velo}\right), \eta_{end}\right] .
\end{equation}

Here $velo = 51$ for all simulations of both CAC-CIM-CDP and OL-CIM-CDP in artificial random data. For the MRI data, $velo = 31$ and $velo = 11$ were used in OL-CIM-CDP and CAC-CIM-CDP respectively. For synthesised random data (Figs. \ref{SAgraphs}, \ref{ppRMSETarggraphs}, and Supplementary note Fig. \ref{GACSsupport}), the threshold $\eta$ was linearly lowered from $\eta_{init}$ to $\eta_{end}$ as the alternating minimisation proceeds. $\eta_{init}$ and $\eta_{end}$ are adjusted to maximise the performance of those models. On the other hand, for MRI data (Figs. \ref{mriAllgraphs}, \ref{mriImagegraphs64}, \ref{mriImagegraphs128}, Supplementary note Fig. \ref{mri64saturationgraphs} and Supplementary note Fig. \ref{mri128maingraphs}), the threshold $\eta$ was constant by setting as $\eta_{init} = \eta_{end}$. The values of $\eta_{init}$ and $\eta_{end}$ used for each simulation are shown in the figure captions.

In both Wigner and P-P type CAC-CIM-CDPs, the target amplitude of CAC, $\tau$, was constant with respect to the time $t$. 
For the simulations in Fig. \ref{ppRMSETarggraphs}a and Fig. \ref{ppRMSETarggraphs}b, $\tau = 0.21$ was used while in Fig. \ref{ppRMSETarggraphs}c and Fig. \ref{ppRMSETarggraphs}d $\tau$ was set to 0.15. For other simulations, $\tau$ was $1$.

\subsection{Observation model for Compressed Sensing}
\label{observationmodel}

The observation model that is the premise of Eq. (\ref{doublel0}) and Eq. (\ref{l0Hamiltonian}) is defined as follows. 

\begin{equation}
\label{observationmodelmatrix}
\begin{bmatrix}
y^{1} \\
y^{2} \\
\vdots\\
y^{M}
\end{bmatrix} = 
\begin{bmatrix}
A_{1}^{1} & A_{2}^{1} & \cdots & A_{N}^{1} \\
A_{1}^{2} & A_{2}^{2} & \cdots & A_{N}^{2} \\
\vdots  & \vdots  & \ddots & \vdots  \\
A_{1}^{M} & A_{2}^{M} & \cdots & A_{N}^{M}
\end{bmatrix}
\begin{bmatrix}
\xi_{1}x_{1} \\
\xi_{2}x_{2} \\
\vdots\\
\xi_{N}x_{N}
\end{bmatrix}
+
\begin{bmatrix}
w_{noise}^{1} \\
w_{noise}^{2} \\
\vdots\\
w_{noise}^{M}
\end{bmatrix} .
\end{equation}

Here, $A \in\mathbb{R}^{N\times M}$ is the observation matrix, $y\in\mathbb{R}^M$ implies the observation signal,  $x\in\mathbb{R}^N$ and $\xi\in(0,1)^N$ are the true source signal and true support, respectively. $w_{noise}\in\mathbb{R}^M$ indicates the observation noise satisfying $\langle w_{noise}^{k}\rangle =0$ and $\langle w_{noise}^{k} w_{noise}^{k'}\rangle =\nu^2 \delta_{kk'}$. $\nu^2$ is the variance of the observation noise.

\subsection{Artificial random data}
\label{randomdata}

To verify the performance of the proposed models statistically and moreover compare those results with ground states predicted with statistical mechanics \cite{Aonishi}, we used many samples of artificial random data $y\in\mathbb{R}^M$ synthesised from the observation model Eq. (\ref{observationmodelmatrix}) in which the values of all entries were randomly determined as follows. Each entry of the observation matrix $A \in\mathbb{R}^{M\times N}$ is randomly generated from an independent and identical normal distribution with the variance of $1/M$, which satisfies $\langle A_r^k\rangle =0$ and $\langle A_r^k A_{r'}^{t'}\rangle = 1/M \delta_{rr'} \delta(kk')$.

Each entry of the true source signal $x\in\mathbb{R}^N$ is randomly generated from an independent and identical normal distribution with the variance of $1$, which satisfies $\langle x_r\rangle = 0$ and $\langle x_r x_{r'}\rangle = \delta_{rr'}$. $a\times N$ elements of $\xi\in(0,1)^N$ are randomly selected and assigned $1$ while others are assigned $0$. $a$ is the sparseness defined in the Introduction.  


\subsection{Simulations with MRI data}
\label{MRIdata}

To evaluate the performance of the proposed models on realistic data, we used MRI data provided from the fastMRI datasets \cite{fastmri}. The initial brain MRI used here was a $320\times 320$ image. To reduce the problem size, we resized the image to $64\times64$ and $128\times128$ images with the BILINEAR interpolation method. We applied the Haar-wavelet transform (HWT) to the two different-sized images and in Fig. \ref{mriImagegraphs64} and Fig. \ref{mriImagegraphs128} we set 78.8\% and 82.2\% of the HWT coefficients to zero to create two different-sized sparse images ($64\times64$ and $128\times128$ pixels) spanned by Haar basis functions with a sparseness of 0.212 and 0.178, respectively. Then, we applied the discrete Fourier transform (DFT) to the two different-sized sparse images to obtain $64\times64$ and $128\times128$ $k$-space data, respectively. Finally, we undersampled 1638 and 4915 points from the $64\times64$ and $128\times128$ $k$-space data at random red points (Fig. \ref{mriImagegraphs64}b and Fig. \ref{mriImagegraphs128}b) to create two observation signals with a compression rate of 0.4 and 0.3 respectively. 

In accordance with our previous work, we sought to reconstruct the source signals from the undersampled $k$-space data by solving the following optimisation problem with CAC-CIM-CDP and OL-CIM-CDP.

\begin{equation}
\label{MRIl0init1}
    x = \operatorname{argmin}(\| y - SFx\|_{2}^{2} + \dfrac{1}{2}\gamma \|\Delta_{v}x\|_{2}^{2} +  \dfrac{1}{2}\gamma \|\Delta_{h}x\|_{2}^{2} + \lambda\|\Psi x\|_{0}) .
\end{equation}

Here, $x$ is a source signal, and $y$ is the observation signal constructed through the above steps. $F$ indicates the DFT matrix and $\Psi$ is the HWT matrix. $F$ and $\Psi$ are orthogonal matrices and their transpose matrices correspond to inverse DFT and inverse HWT, respectively. $S$ is an undersampling matrix executing undersampling at random red points shown in Fig. \ref{mriImagegraphs64}b and Fig. \ref{mriImagegraphs128}b. $\Delta_{v}$ and $\Delta_{h}$ are the matrices discretely representing the vertical and horizontal second-order derivative operators, respectively. $\gamma$ and $\lambda$ are the $l_2$ and $l_0$ regularisation parameters. 

To implement the optimisation problem in Eq. (\ref{MRIl0init1}) on CIM, we estimate the HWT coefficients instead of the pixel values of the image. Applying the HWT $r = \Psi x$ to Eq. (\ref{MRIl0init1}), the mutual interaction matrix $J$ and the Zeeman term vector $h^z$ for CIM are given as

\begin{equation}
\label{MRIzeeman}
    h^{z} = SF\Psi^{T}y,
\end{equation}
\begin{equation}
\label{MRIJMatrix1}
    \tilde{J} = \Psi F^{T}S^{T}SF\Psi^{T} + \gamma\Psi\Delta_{v}^{T}\Delta_{v}\Psi^{T} + \gamma\Psi\Delta_{h}^{T}\Delta_{h}\Psi^{T} .
\end{equation}

Here, the observation matrix is given as $A = SF\Psi^T$. The second and third terms in $\tilde{J}$ are from the $l_2$ regularisation terms. 
After the alternating minimisation, the output of the CDP, $r$, is transformed to the image, $x$, with the inverse HWT $x=\Psi^T r$. $\gamma$ is set to 0.0001. $\tilde{K}$ for OL-CIM-CDP was set to 0.25 while ${K}$ for CAC-CIM-CDP was 0.01. Here we use LASSO's solution as the initial condition for the CIM simulation. 

\backmatter

\bmhead{Authors' contributions}

M.D.S.H.G., and T.A., modelled the system, performed the numerical simulations for the proposed models and wrote the manuscript. M.D.S.H.G., T.A., and S.K., worked on the evaluation of the models. K.M., and M.O., provided feedback on numerical simulations. S.K., Y.I., and Y.Y., helped with the physics of CIM and provided feedback.

\bmhead{Availability of data and materials}

The data generated and/or analysed during this study are not publicly available for legal/ethical reasons. But M.D.S.H.G. can provide the raw data if formally requested.

\bmhead{Competing Interests}

The authors declare no competing interests.

\bmhead{Funding}

This work is supported by the Japan Science and Technology Agency through its ImPACT program, NTT Research Inc. And Authors acknowledges the support of the NSF CIM Expedition award (CCF-1918549).

\bibliography{refs} 

\begin{thebibliography}{10}
\expandafter\ifx\csname url\endcsname\relax
  \def\url#1{\burl{#1}}\fi
\expandafter\ifx\csname urlprefix\endcsname\relax\def\urlprefix{URL }\fi
\providecommand{\bibinfo}[2]{#2}
\providecommand{\eprint}[2][]{\url{#2}}
\providecommand{\doi}[1]{\url{https://doi.org/#1}}
\bibcommenthead

\bibitem{CSastro}
\bibinfo{author}{Bobin, J.}, \bibinfo{author}{Starck, J.-L.} \&
  \bibinfo{author}{Ottensamer, R.}
\newblock \bibinfo{title}{Compressed sensing in astronomy}.
\newblock \emph{\bibinfo{journal}{IEEE Journal of Selected Topics in Signal
  Processing}} \textbf{\bibinfo{volume}{2}}~(5), \bibinfo{pages}{718--726}
  (\bibinfo{year}{2008}).
\newblock \doi{10.1109/JSTSP.2008.2005337} .

\bibitem{CSastro2}
\bibinfo{author}{Zhang, Y.}, \bibinfo{author}{Jiang, J.} \&
  \bibinfo{author}{Zhang, G.}
\newblock \bibinfo{title}{Compression of remotely sensed astronomical image
  using wavelet-based compressed sensing in deep space exploration}.
\newblock \emph{\bibinfo{journal}{Remote Sensing}}
  \textbf{\bibinfo{volume}{13}}~(2) (\bibinfo{year}{2021}).
\newblock \urlprefix\url{https://www.mdpi.com/2072-4292/13/2/288}.
\newblock \doi{10.3390/rs13020288} .

\bibitem{CSastro3}
\bibinfo{author}{Zhang, Y.}, \bibinfo{author}{Jiang, J.} \&
  \bibinfo{author}{Zhang, G.}
\newblock \bibinfo{title}{Compression of remotely sensed astronomical image
  using wavelet-based compressed sensing in deep space exploration}.
\newblock \emph{\bibinfo{journal}{Remote Sensing}}
  \textbf{\bibinfo{volume}{13}}~(2) (\bibinfo{year}{2021}).
\newblock \urlprefix\url{https://www.mdpi.com/2072-4292/13/2/288}.
\newblock \doi{10.3390/rs13020288} .

\bibitem{CSastro4}
\bibinfo{author}{Zhou, W.-P.}, \bibinfo{author}{Li, Y.}, \bibinfo{author}{Liu,
  Q.-S.}, \bibinfo{author}{Wang, G.-D.} \& \bibinfo{author}{Liu, Y.}
\newblock \bibinfo{title}{Fast compression and reconstruction of astronomical
  images based on compressed sensing}.
\newblock \emph{\bibinfo{journal}{Research in Astronomy and Astrophysics}}
  \textbf{\bibinfo{volume}{14}}~(9), \bibinfo{pages}{1207}
  (\bibinfo{year}{2014}) .

\bibitem{CSsig2}
\bibinfo{author}{Herman, M.~A.} \& \bibinfo{author}{Strohmer, T.}
\newblock \bibinfo{title}{High-resolution radar via compressed sensing}.
\newblock \emph{\bibinfo{journal}{IEEE Transactions on Signal Processing}}
  \textbf{\bibinfo{volume}{57}}~(6), \bibinfo{pages}{2275--2284}
  (\bibinfo{year}{2009}).
\newblock \doi{10.1109/TSP.2009.2014277} .

\bibitem{CSbiosig}
\bibinfo{author}{Mamaghanian, H.}, \bibinfo{author}{Khaled, N.},
  \bibinfo{author}{Atienza, D.} \& \bibinfo{author}{Vandergheynst, P.}
\newblock \bibinfo{title}{Compressed sensing for real-time energy-efficient ecg
  compression on wireless body sensor nodes}.
\newblock \emph{\bibinfo{journal}{IEEE Transactions on Biomedical Engineering}}
  \textbf{\bibinfo{volume}{58}}~(9), \bibinfo{pages}{2456--2466}
  (\bibinfo{year}{2011}).
\newblock \doi{10.1109/TBME.2011.2156795} .

\bibitem{obuchi}
\bibinfo{author}{Obuchi, T.}, \bibinfo{author}{Nakanishi-Ohno, Y.},
  \bibinfo{author}{Okada, M.} \& \bibinfo{author}{Kabashima, Y.}
\newblock \bibinfo{title}{Statistical mechanical analysis of sparse linear
  regression as a variable selection problem}.
\newblock \emph{\bibinfo{journal}{Journal of Statistical Mechanics: Theory and
  Experiment}} \textbf{\bibinfo{volume}{2018}}~(10), \bibinfo{pages}{103401}
  (\bibinfo{year}{2018}).
\newblock \urlprefix\url{https://doi.org/10.1088/1742-5468/aae02c}.
\newblock \doi{10.1088/1742-5468/aae02c} .

\bibitem{kabashima}
\bibinfo{author}{Kabashima, Y.}, \bibinfo{author}{Wadayama, T.} \&
  \bibinfo{author}{Tanaka, T.}
\newblock \bibinfo{title}{A typical reconstruction limit for compressed sensing
  based on $l_p$-norm minimization}.
\newblock \emph{\bibinfo{journal}{Journal of Statistical Mechanics: Theory and
  Experiment}} \textbf{\bibinfo{volume}{2009}}~(09), \bibinfo{pages}{L09003}
  (\bibinfo{year}{2009}).
\newblock \urlprefix\url{https://doi.org/10.1088/1742-5468/2009/09/l09003}.
\newblock \doi{10.1088/1742-5468/2009/09/l09003} .

\bibitem{twofold1}
\bibinfo{author}{Louizos, C.}, \bibinfo{author}{Welling, M.} \&
  \bibinfo{author}{Kingma, D.~P.}
\newblock \bibinfo{title}{Learning sparse neural networks through $l_0$
  regularization} (\bibinfo{year}{2017}).
\newblock \urlprefix\url{https://arxiv.org/abs/1712.01312}.

\bibitem{twofold2}
\bibinfo{author}{Nakanishi-Ohno, Y.}, \bibinfo{author}{Obuchi, T.},
  \bibinfo{author}{Okada, M.} \& \bibinfo{author}{Kabashima, Y.}
\newblock \bibinfo{title}{Sparse approximation based on a random overcomplete
  basis}.
\newblock \emph{\bibinfo{journal}{Journal of Statistical Mechanics: Theory and
  Experiment}} \textbf{\bibinfo{volume}{2016}}~(6), \bibinfo{pages}{063302}
  (\bibinfo{year}{2016}).
\newblock \urlprefix\url{https://dx.doi.org/10.1088/1742-5468/2016/06/063302}.
\newblock \doi{10.1088/1742-5468/2016/06/063302} .

\bibitem{Aonishi}
\bibinfo{author}{Aonishi, T.}, \bibinfo{author}{Mimura, K.},
  \bibinfo{author}{Okada, M.} \& \bibinfo{author}{Yamamoto, Y.}
\newblock \bibinfo{title}{L0 regularization-based compressed sensing with
  quantum-classical hybrid approach} (\bibinfo{year}{2021}).
\newblock \eprint{2102.11412}.

\bibitem{cimqubo1}
\bibinfo{author}{Mohseni, N.}, \bibinfo{author}{McMahon, P.~L.} \&
  \bibinfo{author}{Byrnes, T.}
\newblock \bibinfo{title}{Ising machines as hardware solvers of combinatorial
  optimization problems}.
\newblock \emph{\bibinfo{journal}{Nature Reviews Physics}}
  \textbf{\bibinfo{volume}{4}}~(6), \bibinfo{pages}{363--379}
  (\bibinfo{year}{2022}) .

\bibitem{cimqubo2}
\bibinfo{author}{Matsumoto, N.}, \bibinfo{author}{Hamakawa, Y.},
  \bibinfo{author}{Tatsumura, K.} \& \bibinfo{author}{Kudo, K.}
\newblock \bibinfo{title}{Distance-based clustering using qubo formulations}.
\newblock \emph{\bibinfo{journal}{Scientific reports}}
  \textbf{\bibinfo{volume}{12}}~(1), \bibinfo{pages}{1--10}
  (\bibinfo{year}{2022}) .

\bibitem{cimqubo3}
\bibinfo{author}{Tanahashi, K.}, \bibinfo{author}{Takayanagi, S.},
  \bibinfo{author}{Motohashi, T.} \& \bibinfo{author}{Tanaka, S.}
\newblock \bibinfo{title}{Application of ising machines and a software
  development for ising machines}.
\newblock \emph{\bibinfo{journal}{Journal of the Physical Society of Japan}}
  \textbf{\bibinfo{volume}{88}}~(6), \bibinfo{pages}{061010}
  (\bibinfo{year}{2019}).
\newblock \urlprefix\url{https://doi.org/10.7566/JPSJ.88.061010}.
\newblock \doi{10.7566/JPSJ.88.061010},
  \bibinfo{eprint}{{\href{https://arxiv.org/abs/https://doi.org/10.7566/JPSJ.88.061010}{{https://doi.org/10.7566/JPSJ.88.061010}}}}
  .

\bibitem{AonishiCDMA}
\bibinfo{author}{Aonishi, T.}, \bibinfo{author}{Mimura, K.},
  \bibinfo{author}{Okada, M.} \& \bibinfo{author}{Yamamoto, Y.}
\newblock \bibinfo{title}{Statistical mechanics of cdma multiuser detector
  implemented in coherent ising machine}.
\newblock \emph{\bibinfo{journal}{Journal of Applied Physics}}
  \textbf{\bibinfo{volume}{124}}~(23), \bibinfo{pages}{233102}
  (\bibinfo{year}{2018}).
\newblock \urlprefix\url{https://doi.org/10.1063/1.5041998}.
\newblock \doi{10.1063/1.5041998},
  \bibinfo{eprint}{{\href{https://arxiv.org/abs/https://doi.org/10.1063/1.5041998}{{https://doi.org/10.1063/1.5041998}}}}
  .

\bibitem{kako}
\bibinfo{author}{Kako, S.} \emph{et~al.}
\newblock \bibinfo{title}{Coherent ising machines with error correction
  feedback}.
\newblock \emph{\bibinfo{journal}{Advanced Quantum Technologies}}
  \textbf{\bibinfo{volume}{3}}~(11), \bibinfo{pages}{2000045}
  (\bibinfo{year}{2020}).
\newblock
  \urlprefix\url{https://onlinelibrary.wiley.com/doi/abs/10.1002/qute.202000045}.
\newblock \doi{https://doi.org/10.1002/qute.202000045},
  \bibinfo{eprint}{{\href{https://arxiv.org/abs/https://onlinelibrary.wiley.com/doi/pdf/10.1002/qute.202000045}{{https://onlinelibrary.wiley.com/doi/pdf/10.1002/qute.202000045}}}}
  .

\bibitem{Inui2022}
\bibinfo{author}{Inui, Y.}, \bibinfo{author}{Gunathilaka, M. D. S.~H.},
  \bibinfo{author}{Kako, S.}, \bibinfo{author}{Aonishi, T.} \&
  \bibinfo{author}{Yamamoto, Y.}
\newblock \bibinfo{title}{Control of amplitude homogeneity in coherent ising
  machines with artificial zeeman terms}.
\newblock \emph{\bibinfo{journal}{Communications Physics}}
  \textbf{\bibinfo{volume}{5}}~(1), \bibinfo{pages}{154}
  (\bibinfo{year}{2022}).
\newblock \urlprefix\url{https://doi.org/10.1038/s42005-022-00927-x}.
\newblock \doi{10.1038/s42005-022-00927-x} .

\bibitem{AmpLeleu}
\bibinfo{author}{Leleu, T.}, \bibinfo{author}{Yamamoto, Y.},
  \bibinfo{author}{McMahon, P.~L.} \& \bibinfo{author}{Aihara, K.}
\newblock \bibinfo{title}{Destabilization of local minima in analog spin
  systems by correction of amplitude heterogeneity}.
\newblock \emph{\bibinfo{journal}{Phys. Rev. Lett.}}
  \textbf{\bibinfo{volume}{122}}, \bibinfo{pages}{040607}
  (\bibinfo{year}{2019}).
\newblock
  \urlprefix\url{https://link.aps.org/doi/10.1103/PhysRevLett.122.040607}.
\newblock \doi{10.1103/PhysRevLett.122.040607} .

\bibitem{Leleu1}
\bibinfo{author}{Leleu, T.}, \bibinfo{author}{Yamamoto, Y.},
  \bibinfo{author}{Utsunomiya, S.} \& \bibinfo{author}{Aihara, K.}
\newblock \bibinfo{title}{Combinatorial optimization using dynamical phase
  transitions in driven-dissipative systems}.
\newblock \emph{\bibinfo{journal}{Phys. Rev. E}} \textbf{\bibinfo{volume}{95}},
  \bibinfo{pages}{022118} (\bibinfo{year}{2017}).
\newblock \urlprefix\url{https://link.aps.org/doi/10.1103/PhysRevE.95.022118}.
\newblock \doi{10.1103/PhysRevE.95.022118} .

\bibitem{sam}
\bibinfo{author}{Reifenstein, S.}, \bibinfo{author}{Kako, S.},
  \bibinfo{author}{Khoyratee, F.}, \bibinfo{author}{Leleu, T.} \&
  \bibinfo{author}{Yamamoto, Y.}
\newblock \bibinfo{title}{Coherent ising machines with optical error correction
  circuits}.
\newblock \emph{\bibinfo{journal}{Advanced Quantum Technologies}}
  \textbf{\bibinfo{volume}{4}}~(11), \bibinfo{pages}{2100077}
  (\bibinfo{year}{2021}).
\newblock
  \urlprefix\url{https://onlinelibrary.wiley.com/doi/abs/10.1002/qute.202100077}.
\newblock \doi{https://doi.org/10.1002/qute.202100077},
  \bibinfo{eprint}{{\href{https://arxiv.org/abs/https://onlinelibrary.wiley.com/doi/pdf/10.1002/qute.202100077}{{https://onlinelibrary.wiley.com/doi/pdf/10.1002/qute.202100077}}}}
  .

\bibitem{disSimBif}
\bibinfo{author}{Goto, H.} \emph{et~al.}
\newblock \bibinfo{title}{High-performance combinatorial optimization based on
  classical mechanics}.
\newblock \emph{\bibinfo{journal}{Science Advances}}
  \textbf{\bibinfo{volume}{7}}~(6), \bibinfo{pages}{eabe7953}
  (\bibinfo{year}{2021}).
\newblock
  \urlprefix\url{https://www.science.org/doi/abs/10.1126/sciadv.abe7953}.
\newblock \doi{10.1126/sciadv.abe7953},
  \bibinfo{eprint}{{\href{https://arxiv.org/abs/https://www.science.org/doi/pdf/10.1126/sciadv.abe7953}{{https://www.science.org/doi/pdf/10.1126/sciadv.abe7953}}}}
  .

\bibitem{Tibshirani}
\bibinfo{author}{Tibshirani, R.}
\newblock \bibinfo{title}{Regression shrinkage and selection via the lasso}.
\newblock \emph{\bibinfo{journal}{Journal of the Royal Statistical Society.
  Series B (Methodological)}} \textbf{\bibinfo{volume}{58}}~(1),
  \bibinfo{pages}{267--288} (\bibinfo{year}{1996}).
\newblock \urlprefix\url{http://www.jstor.org/stable/2346178} .

\bibitem{npg}
\bibinfo{author}{Ng, E.} \emph{et~al.}
\newblock \bibinfo{title}{Efficient sampling of ground and low-energy ising
  spin configurations with a coherent ising machine}.
\newblock \emph{\bibinfo{journal}{Phys. Rev. Research}}
  \textbf{\bibinfo{volume}{4}}, \bibinfo{pages}{013009} (\bibinfo{year}{2022}).
\newblock
  \urlprefix\url{https://link.aps.org/doi/10.1103/PhysRevResearch.4.013009}.
\newblock \doi{10.1103/PhysRevResearch.4.013009} .

\bibitem{nonoiseCIM}
\bibinfo{author}{Wang, Z.}, \bibinfo{author}{Marandi, A.},
  \bibinfo{author}{Wen, K.}, \bibinfo{author}{Byer, R.~L.} \&
  \bibinfo{author}{Yamamoto, Y.}
\newblock \bibinfo{title}{Coherent ising machine based on degenerate optical
  parametric oscillators}.
\newblock \emph{\bibinfo{journal}{Phys. Rev. A}} \textbf{\bibinfo{volume}{88}},
  \bibinfo{pages}{063853} (\bibinfo{year}{2013}).
\newblock \urlprefix\url{https://link.aps.org/doi/10.1103/PhysRevA.88.063853}.
\newblock \doi{10.1103/PhysRevA.88.063853} .

\bibitem{Inui}
\bibinfo{author}{Inui, Y.} \& \bibinfo{author}{Yamamoto, Y.}
\newblock \bibinfo{title}{Noise correlation and success probability in coherent
  ising machines} (\bibinfo{year}{2020}).
\newblock \eprint{2009.10328}.

\bibitem{drumOps}
\bibinfo{author}{Drummond, P.~D.} \& \bibinfo{author}{Gardiner, C.~W.}
\newblock \bibinfo{title}{Generalised p-representations in quantum optics}.
\newblock \emph{\bibinfo{journal}{Journal of Physics A: Mathematical and
  General}} \textbf{\bibinfo{volume}{13}}~(7), \bibinfo{pages}{2353}
  (\bibinfo{year}{1980}).
\newblock \urlprefix\url{https://dx.doi.org/10.1088/0305-4470/13/7/018}.
\newblock \doi{10.1088/0305-4470/13/7/018} .

\bibitem{fastmri}
\bibinfo{author}{Zbontar, J.} \emph{et~al.}
\newblock \bibinfo{title}{fastmri: An open dataset and benchmarks for
  accelerated mri} (\bibinfo{year}{2018}).
\newblock \urlprefix\url{https://arxiv.org/abs/1811.08839}.

\end{thebibliography}

\end{document}


\maketitle


\section*{Supplementary Note 1: Performance dependence on system size}
\label{systemsize}

\begin{figure}[H]
\centering
\includegraphics[scale=0.33]{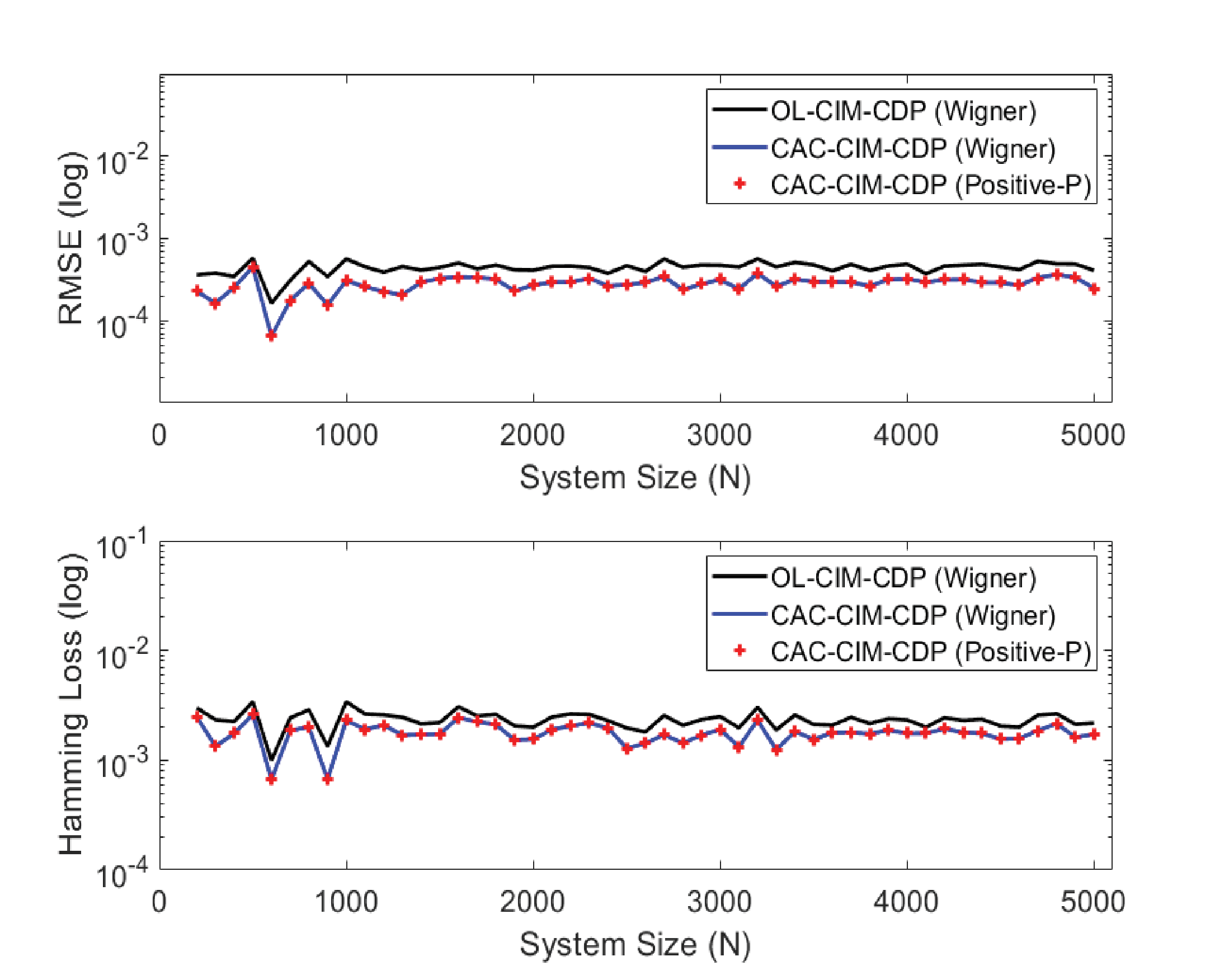}
\caption{\textbf{Performance on different System Sizes}}
\medskip
\justifying
{\textbf{(a)} Log average RMSE values for 10 random Hamiltonians for each system size. For CAC-CIM-CDP (Wigner and Positive-$P$) $j = 1.0$, and $\beta = 1.0$ were used. $\eta_{init}$ and $\eta_{end}$ assigned as 0.9 and 0.01 respectively. For OL-CIM-CDP $\eta_{init}$ and $\eta_{end}$ were set as 0.6 and 0.01 respectively. \textbf{(b)} Log average Hamming loss values for 10 random Hamiltonians for each system size calculated according to Eq. (\ref{hamming}) which calculates the average support estimation accuracy.}
\label{GACSsupport}
\end{figure}

This section compares how the performance of each model differs as the system size increases. In this case, $N$ is in the range of [200, 5000], with compression $\alpha = 0.6$ and sparseness $a = 0.2$. Fig. \ref{GACSsupport} illustrates the acquired results. Black, blue and red marks indicate the log average RMSE values acquired for 10 random hamiltonian simulations for OL-CIM-CDP, CAC-CIM-CDP (Wigner) and CAC-CIM-CDP (Positive-$P$) respectively.

As a measure of model performance, we have used average RMSE (Root Mean Squared Error) and average Hamming loss of support estimation from CIM. Using the following formula (\ref{hamming}), we calculated the hamming loss.

    \begin{equation}
    \label{hamming}
        Hamming \ Loss \ = \dfrac{1}{N} \left( \sum_{r = 0}^{N} \lvert\sigma_r^{CIM}- \xi_r\rvert \right) .
    \end{equation}
    
Here $\xi_{r}$ corresponds to the correct support while $\sigma_{r}^{CIM}$ is the estimation produced by CIM models.
Based on the RMSE values in the top graph of Fig. \ref{GACSsupport}, we can state that even when the size of the system changes, CAC-CIM-CDP models maintain a lower average RMSE value than OL-CIM-CDP.
In addition, the corresponding hamming loss values (bottom graph of Fig. \ref{GACSsupport}) are consistent also with the RMSE results. This suggests that CAC-CIM-CDP has performed a more accurate sparse reconstruction.

Thus, we can conclude that CAC-CIM-CDP provides better results than OL-CIM-CDP, regardless of system size.

\section*{Supplementary Note 2: Performance difference on relatively harder problem instances}
\label{harderist}

Here we discuss the observation we had on the amplitude evolution of CAC-CIM-CDP models when the ${a}/{\alpha}$ ratio is higher (after exceeding the critical threshold). 

\begin{figure}[H]
\centering
\includegraphics[scale=0.32]{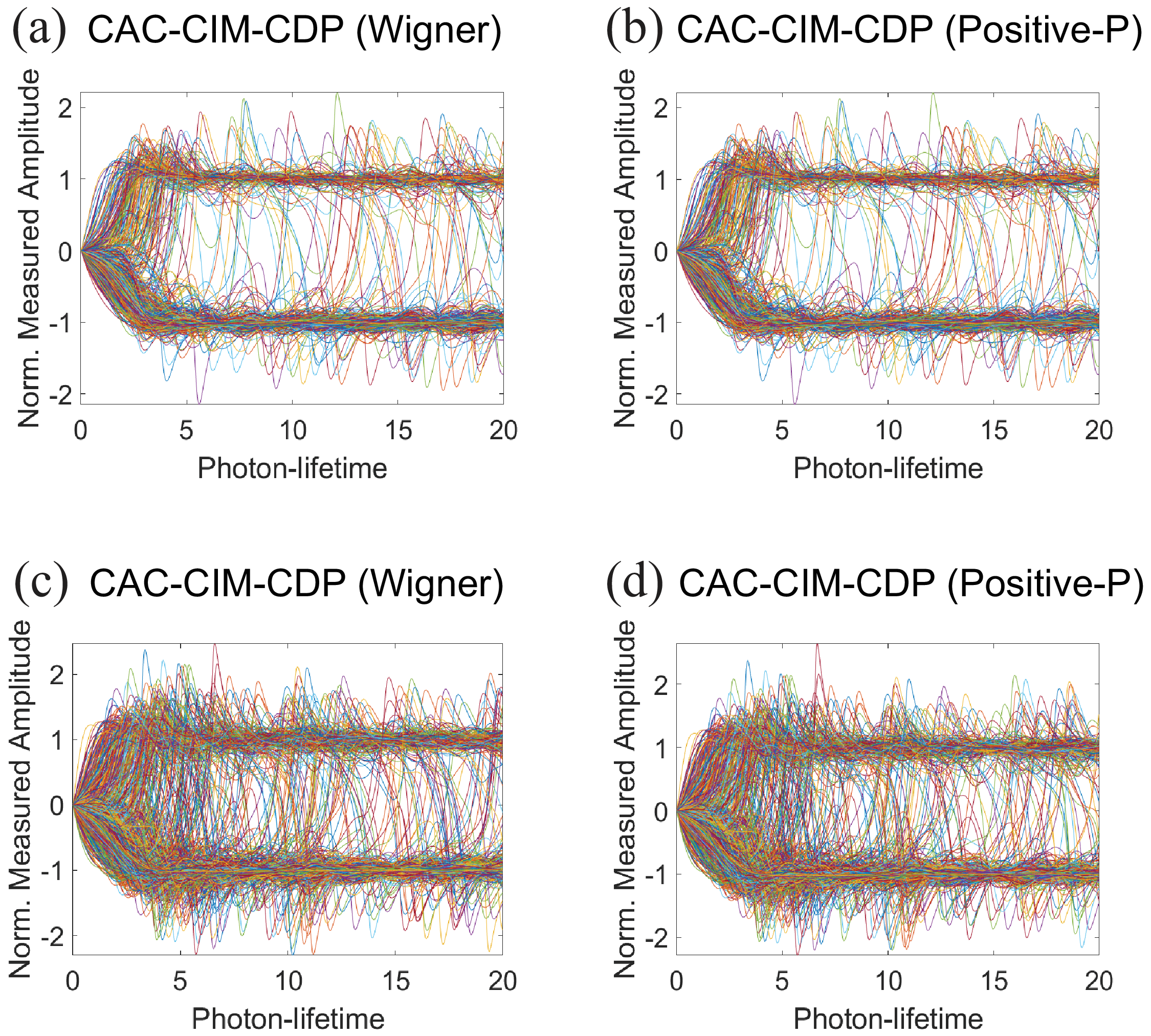}
\caption{\textbf{Normalised measured amplitude $g\tilde{\mu}_r$ evolution for easy and harder problem instances}}
\centering
{\begin{justify}\textbf{(a)} and \textbf{(b)} indicates the normalised amplitude $g\tilde{\mu_r}$ evolution of CAC-CIM-CDP (Wigner and Positive-$P$) with $a = 0.2$ and $\alpha = 0.6$. \textbf{(c)} and \textbf{(d)} indicates the normalised amplitude $g\tilde{\mu}_r$ evolution of CAC-CIM-CDP (Wigner and Positive-$P$) with $a = 0.4$ and $\alpha = 0.6$.\end{justify}}
\label{Nonoiseamplitudes}
\end{figure}

\begin{figure}[H]
\centering
\includegraphics[scale=0.32]{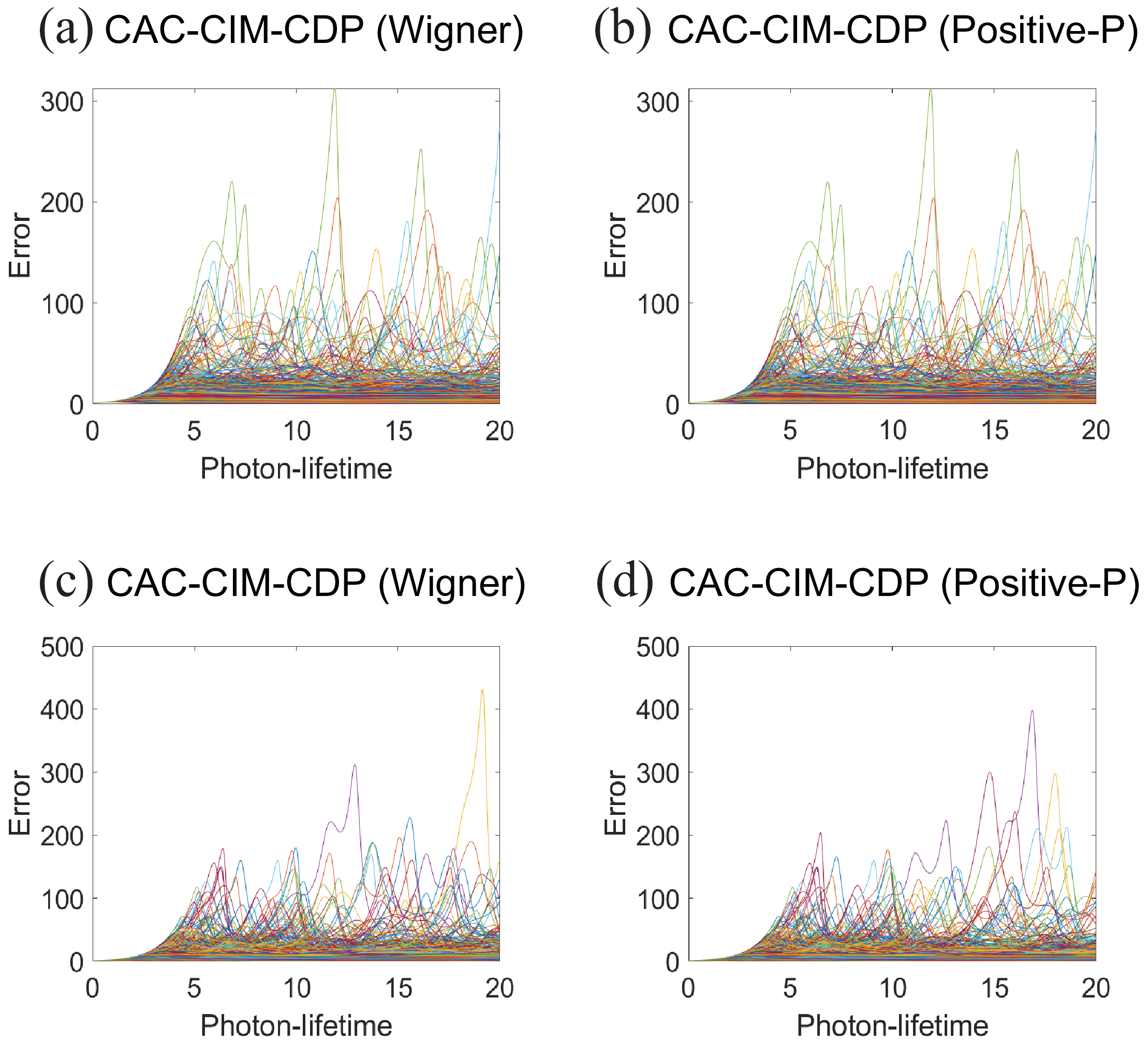}
\caption{\textbf{Error $e_r$ evolution for easy and harder problem instances}}
\centering
{\begin{justify}\textbf{(a)} and \textbf{(b)} indicates the error amplitude $e_r$ evolution of CAC-CIM-CDP (Wigner and Positive-$P$) with $a = 0.2$ and $\alpha = 0.6$. \textbf{(c)} and \textbf{(d)} indicates the error amplitude $e_r$ evolution of CAC-CIM-CDP (Wigner and Positive-$P$) with $a = 0.4$ and $\alpha = 0.6$.\end{justify}}
\label{Nonoiseerrors}
\end{figure}

The difference in performance is clear in normalised measured-amplitude $g\tilde{\mu}_r$ and error $e_r$ as well. In Fig. \ref{Nonoiseamplitudes} and Fig. \ref{Nonoiseerrors} $N = 2000$ system is considered with $\alpha = 0.6$. Fig. \ref{Nonoiseamplitudes}a and Fig. \ref{Nonoiseamplitudes}b $a = 0.2$ is used while Fig. \ref{Nonoiseamplitudes}c and Fig. \ref{Nonoiseamplitudes}d use $a = 0.4$. It is clear that in the $a = 0.4$ case, the amplitude evolution is more chaotic compared to $a = 0.2$ in both models. The corresponding $e_r$ evolution is illustrated in Fig. \ref{Nonoiseerrors}. Even though $a = 0.2$ has identical evolution for CAC-CIM-CDP, with $a = 0.4$ the evolution is slightly different. 

\section*{Supplementary Note 3: Performance with larger saturation parameters}
\label{largerg}

\begin{figure}[H]
\centering
\includegraphics[scale=0.33]{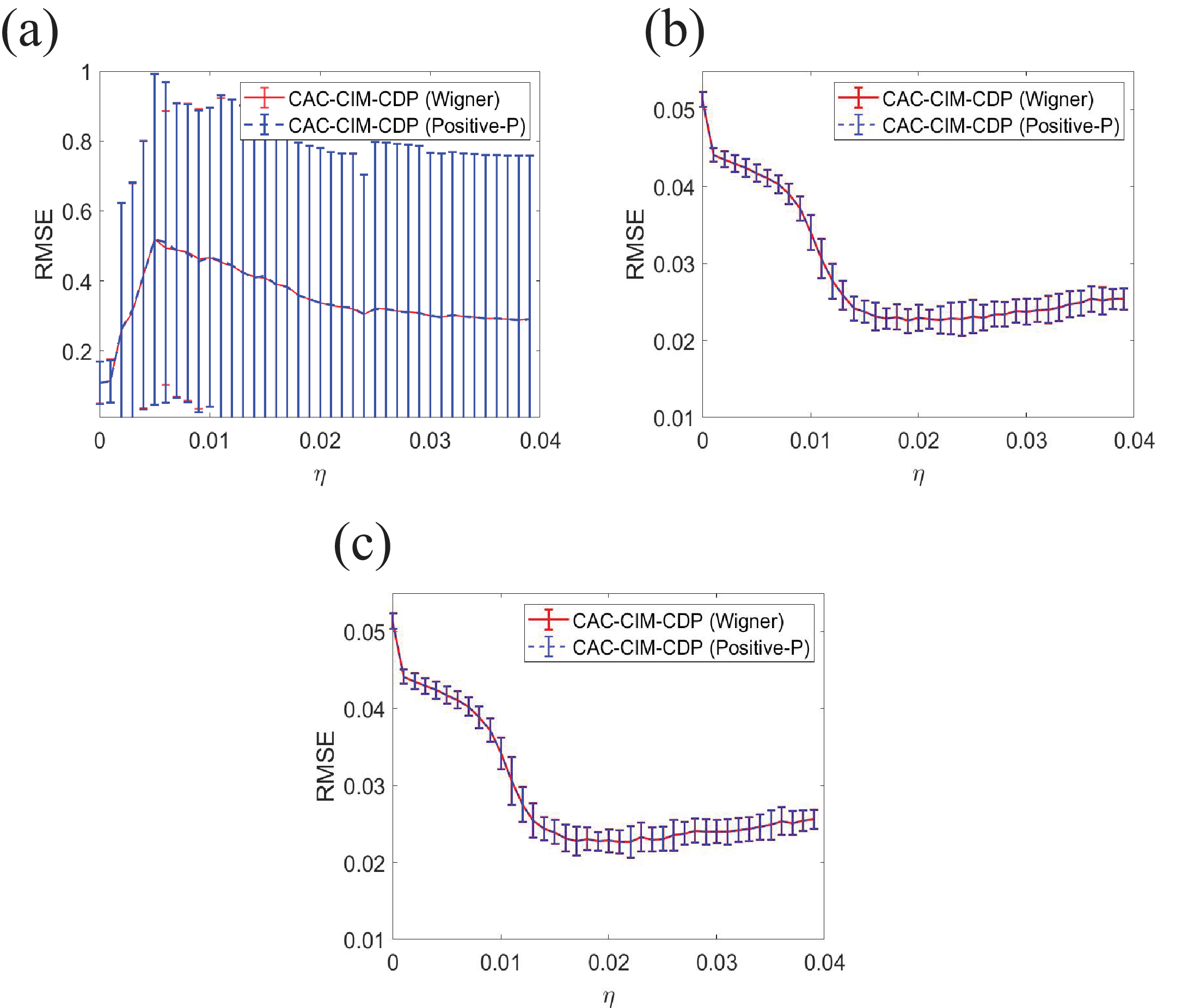}
\caption{\textbf{Performance of CAC-CIM-CDP models with large quantum noise present}}
\centering
{\begin{justify}\textbf{(a)}, \textbf{(b)}, and \textbf{(c)} illustrates the performance of CAC-CIM-CDP models when $g^2$ is $10^{-2}$, $10^{-3}$ and $10^{-7}$ respectively for $64\times 64$ image reconstruction.\end{justify}}
\label{mri64saturationgraphs}
\end{figure}

With the same parameters used for CAC-CIM-CDP (Wigner and Positive-$P$), here we conducted an experiment on how the performance varies when the saturation parameter $g^2$ changes. Parameter $g^2$ corresponds directly to the quantum noise present in the CIM system. The higher the $g^2$, the more quantum noise is present. In the MRI experiments mentioned previously, $g^2 = 10^{-7}$ was used. Which is quite small. In Fig. \ref{mri64saturationgraphs}, we compare CIMs performance with $g^2 = 10^{-2}$, $g^2 = 10^{-3}$, and $g^2 = 10^{-7}$. Red lines indicate the CAC-CIM-CDP (Wigner) and the blue dashed line corresponds to the CAC-CIM-CDP (Positive-$P$) model. Vertical lines are the error bars for each threshold $\eta$ value.

Results indicate that both CAC-CIM-CDP models tend to perform similarly in all the cases. Yet when $g^2 = 10^{-2}$, reconstruction fails. Compared to other $g^2$ values (which have better reconstruction), it implies that for CIM performance there is a threshold saturation parameter value.  

\section*{Supplementary Note 4: Performance change in sparseness}
\label{sparseness}

Here we consider different sparseness $a$ values. For the simulations $64\times 64$ images were used. 

\begin{figure}[H]
\centering
\includegraphics[scale=0.32]{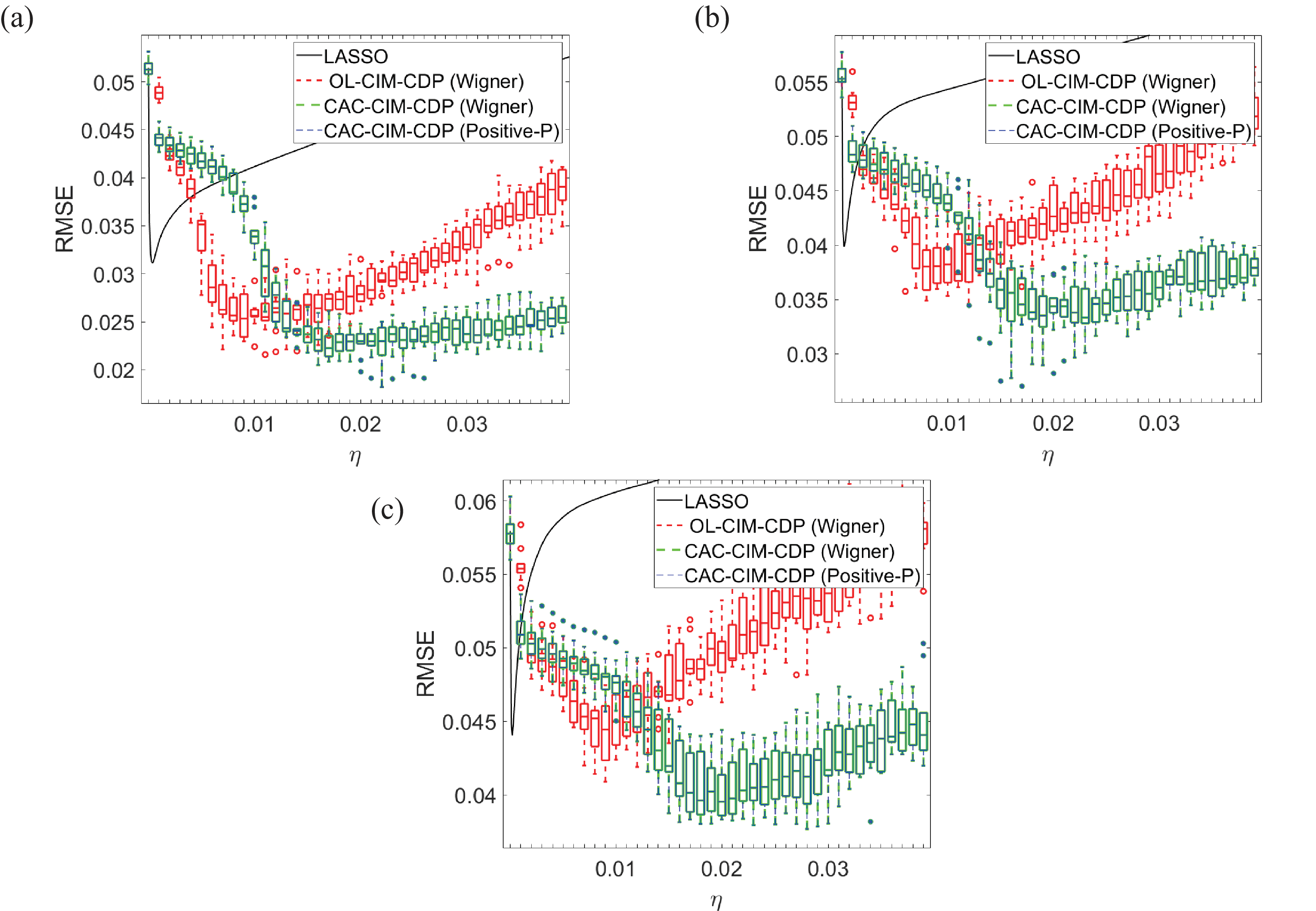}
\caption{\textbf{{Average performance of LASSO, OL-CIM-CDP, CAC-CIM-CDP (Wigner) and CAC-CIM-CDP (Positive-$P$) models when sparseness varies for $64\times64$ MRI image}}}
\centering
\begin{justify}
\textbf{(a)}, \textbf{(b)}, and \textbf{(c)} correspond to the reconstructions obtained from LASSO, OL-CIM-CDP, and CAC-CIM-CDP (Positive-$P$) with sparseness values 0.212, 
0.255, 
and 0.280 
respectively. For every model compression ratio was 0.4. The black line indicates the performance on LASSO while the red boxes correspond to OL-CIM-CDP. Green and blue boxes indicate the performance on CAC-CIM-CDP Wigner and Positive-$P$ respectively. For different threshold values, 
the graphs illustrate the maximum, minimum, 25-th percentile (bottom edge), 75-th percentile (top edge), and median (central horizontal line) of RMSEs for each model with box plots. The markers indicate the outliers.
\end{justify}
\label{mri128maingraphs}
\end{figure}

Reconstruction becomes harder when the $a/\alpha$ ratio becomes larger. In Fig. \ref{mri128maingraphs}, we showcase the results on how the performance of LASSO, OL-CIM-CDP and the two CAC-CIM-CDP models change when $a$ is increased for a constant $\alpha$ value with respect to $\eta$. The green boxplot indicates the CAC-CIM-CDP (Wigner) and the blue boxplot corresponds to the CAC-CIM-CDP (Positive-$P$) model. The black line and red boxplot indicate the LASSO algorithm and OL-CIM-CDP respectively. Based on different threshold values, box plots illustrate the maximum, minimum, 25th percentile (bottom edge), 75th percentile (top edge), and median (central horizontal line) of RMSEs for each model. In the markers, outliers are identified.

With the above results, it is clear that CAC-CIM-CDP models tend to perform better than OL-CIM-CDP. However, when $a$ increases the reconstruction accuracy worsens in all three CIM-CDP models and LASSO. Even though the accuracy deteriorates, CAC-CIM-CDP models still perform better than OL-CIM-CDP. This indicates that even though the problem becomes more challenging, CAC-CIM-CDP is still a better CIM model than the OL-CIM-CDP for $l_0$-regularised CS. 


